\documentclass[journal]{IEEEtran}

\usepackage{multirow}
\usepackage[table,xcdraw]{xcolor}
\usepackage{graphicx}
\usepackage{amsmath}
\usepackage{amsfonts}
\usepackage{algorithm}
\usepackage{amsthm,amssymb}
\usepackage{mathrsfs}
\usepackage{algorithmic}
\usepackage{makecell}
\usepackage{easyReview}
\usepackage{booktabs}
\setcellgapes{3pt}
\usepackage{verbatim}
\setcellgapes{3pt}

\usepackage[justification=raggedright]{caption}
\usepackage{color}
\usepackage[subfigure]{tocloft}
\usepackage{subfigure}
\usepackage{tabularx}
\usepackage{amsmath, bm}
\usepackage{hyperref}
\hypersetup{hidelinks,
	colorlinks=true,
	allcolors=blue,
	pdfstartview=Fit,
	breaklinks=true}

\usepackage{cite}

\graphicspath{{./Attachments/images-pdf/}{./Attachments/savefig.assets}}

\begin{document}

%

\title{Immersive Volumetric Video Playback: Near-RT Resource Allocation and O-RAN-based Implementation}

    \author{Yao Wen,~\IEEEmembership{Graduate~Student~Member,~IEEE,} Luping Xiang,~\IEEEmembership{Senior~Member,~IEEE,} Kun Yang,~\IEEEmembership{Fellow,~IEEE} 


    \IEEEcompsocitemizethanks{
		\IEEEcompsocthanksitem Yao Wen (ywen@smail.nju.edu.cn), Luping Xiang (luping.xiang@nju.edu.cn) and Kun Yang (kunyang@nju.edu.cn) are with State Key Laboratory of Novel Software Technology, Nanjing University, Nanjing, 210008, China, Institute of Intelligent Networks and Communications (NINE), Collaborative Innovation Center of Novel Software Technology and Industrialization, and School of Intelligent Software and Engineering, Nanjing University (Suzhou Campus), Suzhou, 215163, China. 
    }
}

\maketitle


\begin{abstract}
  Immersive volumetric video streaming in extended reality (XR) demands ultra-low motion-to-photon (MTP) latency, which conventional edge-centric architectures struggle to meet due to per-frame computationally intensive rendering tightly coupled with user motion. To address this challenge, we propose an Open Radio Access Network (O-RAN)-integrated playback framework that jointly orchestrates radio, compute, and content resources in near real-time (Near-RT) control loop. The system formulates the content-hit (rendered-pixel) ratio as a continuous control variable and jointly optimizes it over the Open Cloud (O-Cloud) compute, gNB transmit power, and bandwidth under a Weber-Fechner quality of experience (QoE) model, explicitly balancing resolution, computation, and latency. A Soft Actor-Critic (SAC) agent with structured action decomposition and QoE-aware reward shaping resolves the resulting high-dimensional control problem. Experiments on a 5G O-RAN testbed and system simulations show that SAC reduces median MTP latency by more than $11\%$ and improves both mean QoE and fairness, demonstrating the feasibility of RAN Intelligent Controller (RIC)-driven joint radio-compute-content control for scalable, latency-aware immersive streaming.
\end{abstract}


\section{Introduction}

\IEEEPARstart{E}{xtended} Reality (XR) aims to deliver deeply immersive and interactive experiences by seamlessly merging the physical and digital worlds. Its applications already span telemedicine, autonomous driving, education, and entertainment \cite{Amiri2024Application}. XR services, encompassing augmented reality (AR), virtual reality (VR), and cloud gaming (CG) \cite{Gapeyenko2023Standardization}, are typically consumed via head-mounted displays (HMDs) that render three-dimensional (3D) scenes and react to user motion in real time\cite{Chen2023Networking}. The underlying video formats enabling immersion include $360^\circ$/holographic video \cite{Shen2023immersive}, free-viewpoint video (FVV) \cite{Zhang2021Editable}, and immersive volumetric video (ImViD) \cite{Han2020ViVo,Yang2025ImViD}. However, generating novel viewpoints for volumetric content demands per-frame computations tightly coupled with head tracking. Although FVV can synthesize viewpoints by aggregating multiple video streams \cite{Zhang2023Edge-FVV}, such operations incur substantial network load. The combination of massive data volume and stringent motion-to-photon (MTP) latency requirements \cite{Gapeyenko2023Standardization} render XR applications highly delay-sensitive, where excessive MTP can trigger motion sickness \cite{Zhou2025Federated}. Hence, achieving low-latency, high-resolution ImViD delivery is fundamental to maintaining immersive quality of experience (QoE).

Volumetric video playback typically comprises two pipeline stages \cite{Yang2025ImViD}: (i) 3D scene reconstruction, which is computationally intensive but relatively delay-tolerant, and (ii) real-time playback, which is tightly bounded by the inter-frame interval and MTP latency. HMDs, constrained by power and processing resources, struggle to sustain high-quality ImViD rendering. 
To alleviate device burden, numerous studies have explored task offloading to fog or Mobile Edge Computing (MEC) platforms \cite{Liu2019MEC-Assisted,Ojaghi2023SO-RAN,Sun2025Energy,Yu2023QoE}.
Representative examples include F-RAN for distributed computation and caching \cite{You2019Fog}, and ML-based edge allocation schemes for FVV to minimize viewing delay \cite{Zhang2023Edge-FVV}. 
Recent edge human digital twin studies balance accuracy and timeliness via two-timescale optimization under energy and delay constraints \cite{Yang2024Dynamic}. 
Content-aware mechanisms such as viewport/tile streaming and attention-driven delivery \cite{11165352,Zou2021Modeling,Yu2024Attention-Based,Du2023Attention-Aware} further reduce computation and bandwidth consumption by prioritizing perceptually important regions, MEC-assisted VR/AR leveraging field-of-view (FoV) prediction and deep reinforcement learning (DRL) scheduling \cite{Liu2021Learning-Based}, while federated attention prediction enhances QoE under the Weber-Fechner perception model \cite{Zhou2025Federated}. 
Generative-AI-aided digital twin interaction, together with RIS \cite{11139123}, leverages Weber-Fechner QoE and joint communication-computation optimization \cite{Chen2025Generative}.
Despite these advances, MEC nodes are typically farther from users than the Next Generation Node Base (gNB) or O-Cloud, so transport delay can dominate the MTP budget for interactive XR even with queuing-aware offloading \cite{Yi2020Multi-User} and game-theoretic allocation \cite{Cao2018Distributed}.
This motivates integrating communication and computation more tightly within the Radio Access Network (RAN) for the pose-driven per-frame playback loop within stringent MTP deadlines in near real time (Near-RT) scale, and we further enable per-frame MTP control via a continuous content-hit ratio under a Weber-Fechner QoE model.

The Open Radio Access Network (O-RAN) architecture provides a viable framework for such integration. By embracing openness, disaggregation, virtualization, and programmability \cite{Polese2023Understanding}, O-RAN exposes software-defined interfaces for real-time control and reconfiguration \cite{Lacava2024Programmable}. The RAN Intelligent Controller (RIC) serves as a programmable substrate for cross-domain orchestration, allowing third-party xApps/rApps/dApps to coordinate radio and compute resources at multiple time scales \cite{DOro2022dApps,Abdalla2022Next}. Standardized E2 and RAN Control (RC) service models \cite{ORAN-E2SM-KPM-v6.0,ORAN-E2SM-RC-v7.0} provide interfaces to monitor key performance indicators (KPIs) and trigger policy enforcement, while the Service Management and Orchestration (SMO) layer coordinates A1/E2/O2 interactions for network-wide policy realization \cite{ORAN-ORCH-USE-CASES-v13.0}. These capabilities enable dynamic pooling of spectrum, power, and compute resources to jointly optimize end-to-end performance.

\newcommand{\cmark}{$\checkmark$}
\begin{table*}[htbp]
\centering
\caption{Comparison Summary of Major Related Works}
\label{tab:related_work_summary_refined}
\begin{tabular}{p{1.7cm}p{2.8cm}p{4.8cm}p{2cm}p{1.6cm}p{0.5cm}p{1.2cm}}
\hline
Reference & Architecture/Platform & XR Service Objective & Communication & Computation & QoE & Near-RT \\
\hline
\cite{You2019Fog} & F-RAN & Latency, Compute-Storage Tradeoff & \cmark & \cmark &  \\
\hline
\cite{Liu2021Learning-Based} & MEC association & $360^\circ$ video, Interaction Latency, QoE & \cmark (Association) & \cmark & \cmark \\
\hline
\cite{Zou2021Modeling, Han2020ViVo, Du2023Attention-Aware} & MEC & Transmission volume, Bandwidth & \cmark &  &  \\
\hline
\cite{Zhang2023Edge-FVV} & MEC Edge Caching & Reduce Viewing Latency & \cmark &  &  \\
\hline
\cite{Yu2024Attention-Based} & MEC, Digital twin & QoE, Resource Fairness & \cmark (Bandwidth) & \cmark & \cmark \\
\hline
\cite{Chen2025Generative} & Digital twin, RIS & QoE, Beamforming, Phase shift & \cmark (RIS) & \cmark & \cmark \\
\hline
\cite{Bagherinejad2025DRL-Based} & IBS Subnet & Reliable Communication & \cmark & \cmark \\
\hline
\cite{Das2024DeMPUP} & B5G Core Network & Power Consumption, UPF Path & \cmark &  &  \\
\hline
\cite{Li2025Coordinating} & O-RAN, NS3 & End-to-End Latency & \cmark & \cmark &  \\
\hline
\textbf{Our Work} & \textbf{O-RAN, RIC, O-Cloud} & \textbf{ImViD, QoE, Per-frame End-to-End Latency} & \cmark (Bandwidth, Power) & \cmark & \cmark & \cmark \\
\hline
\end{tabular}
\end{table*}

Recent studies have explored O-RAN's potential in supporting latency-critical and immersive services. For instance, coordinated radio-computation control improves predictive offloading and resource orchestration for VR traffic \cite{Li2025Coordinating}, while RIC-driven slicing and power control enhance energy efficiency under service-level agreements (SLAs) \cite{Ojaghi2023SO-RAN,Lima2025Power-Efficient}. Moreover, 3GPP TR 38.838 and NR Releases~17-18 formalize XR traffic characteristics and KPIs, emphasizing cross-layer design that balances throughput, latency, and energy consumption \cite{Gapeyenko2023Standardization,Stoica2024XR-Aware}. Meanwhile, learning-based orchestration has gained traction in O-RAN: DRL algorithms enable sequential spectrum and power allocation \cite{Bagherinejad2025DRL-Based}, federated DRL and prompt-based training improve distributed learning convergence \cite{Yu2024Attention-Based,Singh2024Communication}, and stochastic control approaches (e.g., MAREA) enforce probabilistic latency guarantees \cite{Adamuz-Hinojosa2025MAREA}. 
Open-source implementations such as FlexRIC and xDevSM \cite{Schmidt2021FlexRIC,Feraudo2024xDevSM} further demonstrate the feasibility of RIC-driven orchestration. 
Integration of sensing \cite{10806828} and CUDA acceleration \cite{Villa2025Programmable} into O-RAN further enhances its capability to support real-time applications.
However, most prior works treat radio and compute allocation separately or at coarse timescales, rarely addressing per-frame, user-specific coordination required for immersive volumetric workloads. Table~\ref{tab:related_work_summary_refined} summarizes representative studies by architecture, XR service objectives, and focus dimensions, highlighting that existing MEC- and RIC-based designs lack joint Near-RT orchestration of the O-Cloud compute resources and radio scheduling specifically for ImViD.

\subsection{Challenges}
In summary, O-RAN provides the architectural foundation for joint communication-computation orchestration, yet the problem of per-frame, QoE-aware, Near-RT resource allocation for ImViD playback remains largely unaddressed, as shown in Table~\ref{tab:related_work_summary_refined}. 
Specifically, achieving fine-grained orchestration of gNB bandwidth, transmission power, and the O-Cloud's compute resources within each frame interval poses unique challenges:

\begin{itemize}
\item \textbf{Challenge 1: High rendering cost and strict latency constraints in Near-RT playback.} ImViD rendering scales with pixel density, yet HMDs lack sufficient compute and battery capacity to meet stringent MTP deadlines at Near-RT timescales. Offloading mitigates local load but introduces transport latency, creating an inherent latency-quality tradeoff.
\item \textbf{Challenge 2: Per-frame QoE-aware content control.} Existing methods typically adjust coarse parameters (e.g., frame rate) but neglect fine-grained QoE-aware per-frame content-hit ratio, which more directly captures the balance between perceptual fidelity and computational cost under user experience model.
\item \textbf{Challenge 3: Joint radio-compute orchestration at Near-RT timescales.} Isolated optimization of radio or compute resources ignores their interdependence. Effective ImViD playback requires coordinated Near-RT per-frame scheduling of bandwidth, power, and the O-Cloud compute resources across multiple users scale, to jointly meet MTP and QoE targets.
\item \textbf{Challenge 4: High-dimensional continuous and constrained control space.} Per-user, per-frame control over content-hit ratio, O-Cloud compute allocation, bandwidth, and transmit power yields a high-dimensional continuous action space with hard budget constraints and coupled QoE-latency-fairness objectives, where naive heuristics or conventional RL methods can be unstable or sample-inefficient.
\item \textbf{Challenge 5: Practical deployment and prototype validation limitations.} Constructing a realistic O-RAN testbed for end-to-end validation is costly and time-consuming, and limited access to integrated platforms makes practical deployment and measurement of the proposed system difficult.
\end{itemize}

\subsection{Contributions}

To address the above challenges, we propose an O-RAN-assisted ImViD playback framework that integrates communication and computation resources for low-latency immersive rendering. The main contributions of this work are summarized as follows:
\begin{itemize}
\item We formulate a joint optimization problem coupling the O-Cloud compute resources, gNB power, and bidirectional bandwidth with the Near-RT per-frame content-hit ratio as a controllable variable under a Weber-Fechner QoE model, thereby quantifying the tradeoff among resolution, computation, and delay. (\textbf{Addresses Challenges 1-2})
\item We develop an O-RAN orchestration mechanism for per-user, per-frame coordination between radio scheduling and the O-Cloud's compute resource allocation, incorporating a pixel-to-compute latency model and a practical scheduler for joint optimization, under Near-RT limitation. (\textbf{Addresses Challenge 3})
\item We design a Soft Actor-Critic (SAC)-based reinforcement learning algorithm with structured action decomposition, separating content control from radio-compute allocation, and employ QoE-aware reward shaping aligned with perceptual metrics to enhance training stability. (\textbf{Addresses Challenges 4})
\item We implement the proposed system on a 5G O-RAN prototype and conduct large-scale simulations. Experimental results demonstrate that the proposed approach reduces median MTP latency by above $11\%$ and improves both average QoE, fairness and decision-making efficiency compared with baselines. (\textbf{Addresses Challenges 5})
\end{itemize}

The remainder of this paper is organized as follows. Section~II presents the O-RAN-assisted ImViD playback architecture. Section~III formulates the joint content-hit and radio-compute allocation problem, and details the SAC-based orchestration algorithm. Section~IV details simulation results. Section~V reports prototype validation, followed by conclusions in Section~VI.

\section{O-RAN-Based Volumetric Video Playback Architecture}

This section presents the ImViD playback workflow and its realization on an O-RAN-enabled 5G system. Users consume volumetric content via HMDs, while a gNB-coordinated by the RIC and supported by the O-Cloud, provides tightly integrated communication and computation services.

\subsection{Scenario Overview}
We consider a 5G O-RAN deployment in which HMDs attach to a gNB for ImViD playback. The gNB collaborates with the RIC and the O-Cloud to dynamically orchestrate radio and compute resources, thereby delivering both low-latency connectivity and high-fidelity rendering. The service provider reconstructs a high-quality 3D scene \cite{Yang2025ImViD} by combining multi-view capture, scene reconstruction, and virtual content generation. Consistent with \cite{Xu2024Representing}, the overall pipeline comprises two phases: 3D scene reconstruction and volumetric video playback.

In reconstruction, XR captures multi-view 2D images and reconstructs the 3D representation. Owing to perspective geometry and non-linear inverse rendering, the image-to-geometry mapping is computationally demanding. We adopt 3D Gaussian splatting as a representative rendering primitive and focus our architectural design on the real-time playback phase. Reconstruction is computationally intensive yet comparatively delay-tolerant; large upstream transfers are acceptable, thus, this phase can be executed by the service provider without consuming O-RAN compute resources.

\begin{figure}[htbp]
\centering
\includegraphics[width=1\linewidth]{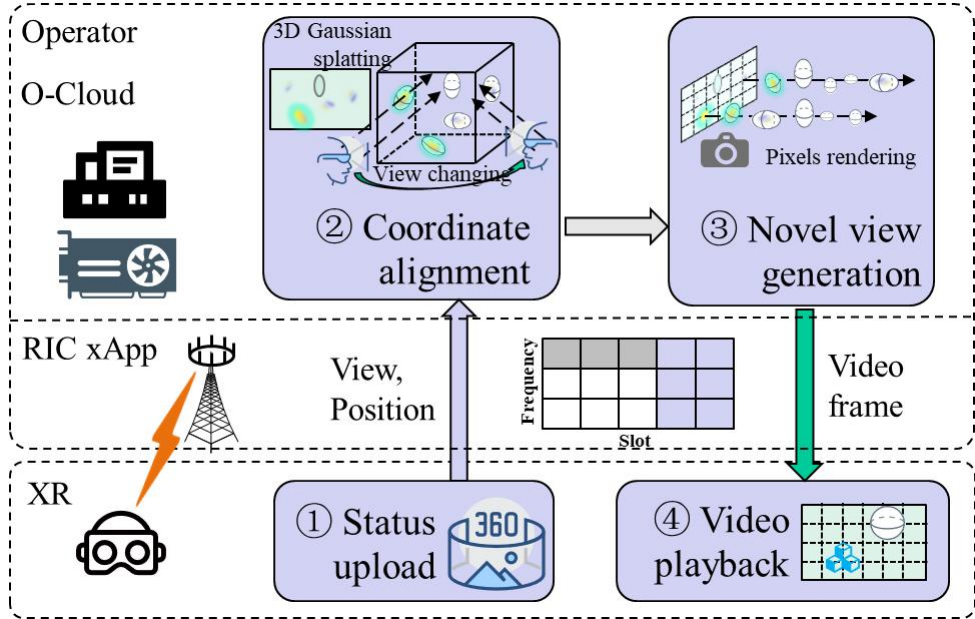}
\caption{ImViD playback workflow per frame.}
\label{fig:imvid_playback_progress}
\end{figure}
During playback, as illustrated in Fig.~\ref{fig:imvid_playback_progress}, each frame proceeds as follows: (1) the HMD reports its instantaneous state (e.g., head pose/position) to an XR xApp on the Near-RT RIC; (2) the xApp computes and forwards the target viewpoint to the O-Cloud for alignment with the 3D model; (3) the O-Cloud renders a novel-view frame; and (4) the xApp schedules download resources and delivers the frame to the HMD. State updates are compact but latency-critical. Rendering complexity scales with resolution because 3D Gaussian splatting represents content as a set of ordered ellipsoids \cite{Kerbl20233D}; pixels are generated independently, so the frame time depends on the rendered-pixel count and the allocated O-Cloud compute resources. The RAN maps the completed frame onto radio resources for over-the-air delivery.

Reconstruction exhibits high upload volume with relaxed latency and mainly affects startup. Playback directly impacts perceived timeliness and comfort: frames must be generated and delivered within the MTP budget. Hence, playback requires joint scheduling of video generation (compute) and video delivery (radio). Offloading generation to the O-Cloud supplies scalable compute and reduces per-frame latency.

\subsection{O-RAN-Assisted ImViD Architecture}

Once the 3D scene is reconstructed, users can explore arbitrary perspectives. Novel-view synthesis and 3D projection are both compute-heavy, while HMDs are constrained by power/battery and cannot meet strict rendering deadlines. Edge offloading alleviates on-device load but may still violate tight MTP budgets. Placing elastic compute in the O-Cloud near the RAN and coordinating it via the RIC provides a practical path to meet frame-level latency targets.

\begin{figure}[!htb]
\centering
\includegraphics[width=1\linewidth]{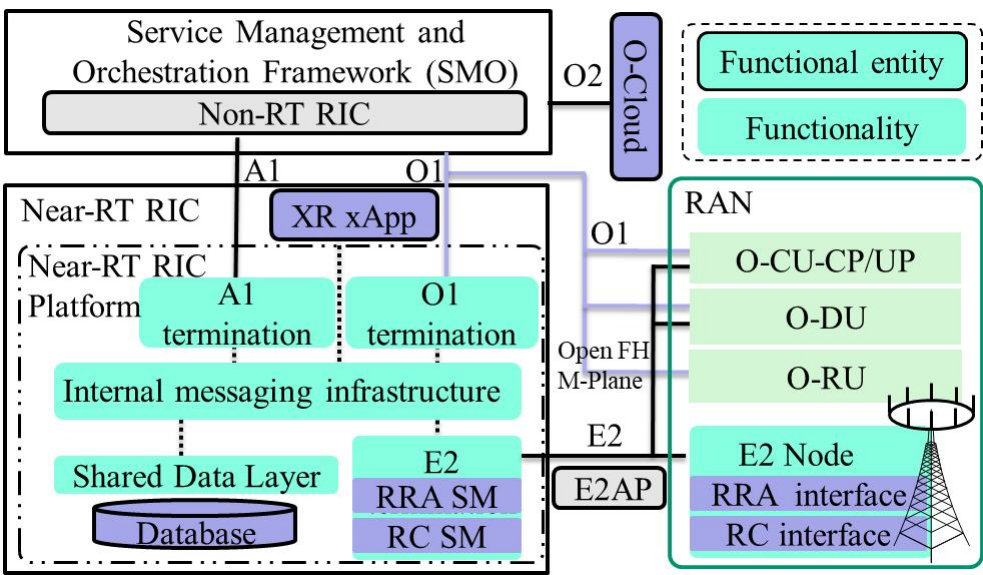}
\caption{O-RAN structural and E2AP control for playback.}
\label{fig:oran_e2ap}
\end{figure}

Our O-RAN-based ImViD system (Fig.~\ref{fig:oran_e2ap}) integrates the SMO, the Non-RT RIC, and the Near-RT RIC to realize coordinated control. The service provider uploads the reconstructed 3D scene to the XR xApp's shared data layer. Upon user request, XR xApp requests the O-Cloud's compute resources via the SMO and instantiates a per-user rendering service. The HMD periodically uploads VR state (e.g., pose) through the gNB; the xApp computes the viewpoint, selects the target pixel set for rendering, and forwards the request to the O-Cloud. After rendering, the frame is mapped to a Data Radio Bearer (DRB) via SDAP at the O\text{-}CU\text{-}UP and transmitted to the HMD. This per-frame closed loop must satisfy the MTP budget.

\subsection{Service Management and Orchestration}

The SMO and Non-RT RIC jointly coordinate the O-Cloud and RAN resources and implement FCAPS (Fault, Configuration, Accounting, Performance, Security) management. As shown in Fig.~\ref{fig:oran_e2ap}, the SMO connects to the O-Cloud via O2 (cloud orchestration) and to the Near-RT RIC via A1 (policy/analytics). It also interfaces with RAN functions and O-RUs through O1 and the Open Fronthaul M-Plane for configuration and performance monitoring. The SMO hosts rApps that interact with the Non-RT RIC over R1 for non-real-time optimization and analytics, enabling end-to-end policy and resource coordination across the O-RAN stack.

\subsection{Non-Real-Time RAN Intelligent Controller}
The Non-RT RIC governs multi-second timescales via rApps, handling policy synthesis, network modeling, slice lifecycle, and infrastructure orchestration. rApps ingest KPIs, demand forecasts, and energy metrics (via O1) and optimize long-term objectives (e.g., latency/energy tradeoffs). Through A1, the Non-RT RIC provides policies (time horizon $>$ 1\,s) and ML model management to the Near-RT RIC; xApps consume these policies for near-real-time decisions. For ImViD, the XR xApp defines per-user compute requirements and interacts with the rApp to provision per-user rendering services in the O-Cloud. Rendered frames produced in the O-Cloud return to the RIC/SMO domain (via O2/O1) and are forwarded by the xApp to the RAN (via E2) for delivery. In practice, SMO/O1/O2/A1 throughput and policy exchange efficiency may bound performance; improving encoding (e.g., ASN.1) and transport (e.g., SCTP) enhances data collection and policy reliability.

\subsection{Near-Real-Time RAN Intelligent Controller}

The Near-RT RIC provides sub-second control ($\sim$10\,ms-1\,s) via xApps that implement radio resource management, power control, mobility, scheduling, and slice control to satisfy ImViD requirements. xApps run as microservices, exchange data via an internal message bus, and access RAN state for decision making. The Near-RT RIC communicates with O-CU/O-DU over the E2 interface, carrying measurement and control messages for near-real-time optimization; xApps collect KPIs and issue spectrum/power commands accordingly. Shared storage and the message infrastructure offer persistent services to the platform and xApps.

FlexRIC \cite{Schmidt2021FlexRIC} is an extensible RIC framework that eases xApp development and deployment. The E2 interface links the Near-RT RIC to O-CU/O-DU and carries E2AP/E2SM messages (over SCTP/IP) for reporting and control. E2AP handles session/service management (setup, indication, reset, service update), while E2SM specifies service-level data formats and logic for different RAN functions.
Developers implement E2 terminations on the Near-RT RIC and corresponding E2 nodes on RAN elements; supported service types include report, insert, control, policy and query. One E2 termination may manage many E2 nodes (one-to-many). xApps use E2 to collect KPIs and to issue spectrum or power control commands. The shared data layer and message infrastructure provide persistent storage and platform data services for xApps.

\begin{figure}[!htb]
\centering
\includegraphics[width=1\linewidth]{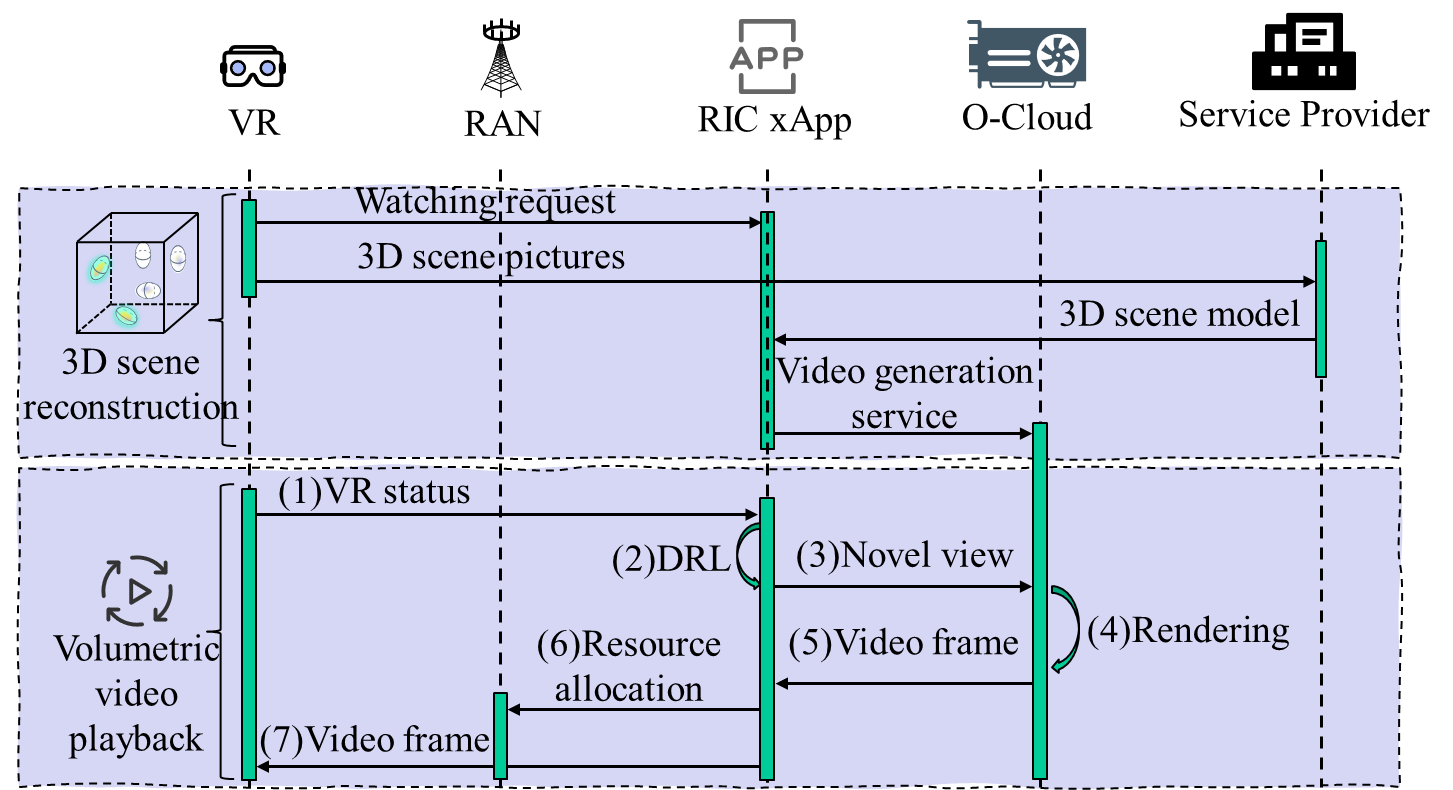}
\caption{Volumetric video reconstruction and playback flow.}
\label{fig:imvid_reconstruction_and_playback}
\end{figure}
As shown in Fig.~\ref{fig:imvid_reconstruction_and_playback}, the XR xApp schedules communication and compute for both phases. During reconstruction, the xApp instantiates services on demand, forwards multi-view uploads to the content provider, and stores the returned 3D model (e.g., from Gaussian splatting) in shared storage. During playback, the HMD sends state updates; the xApp computes the viewpoint and requests rendering from the O-Cloud, which aligns the viewpoint with the stored model, renders the current frame, and returns it for radio delivery. Each loop advances the video by one frame.

\subsection{Roles of O1, O2, A1, and E2 Interfaces}

In an O-RAN-based ImViD system, interfaces have distinct roles. O1 conveys management/telemetry between the SMO and network elements (including the Near-RT RIC and RAN). O2 connects the SMO to the O-Cloud for orchestration. A1 transfers policies and ML artifacts from the Non-RT RIC to the Near-RT RIC (e.g., compute-service allocations, policy identifiers/types/parameters). E2 enables near-real-time control/data exchange between the Near-RT RIC and the RAN. E2 carries, among others, Key Performance Measurements (E2SM-KPM) \cite{ORAN-E2SM-KPM-v6.0} and RAN Control (E2SM-RC) \cite{ORAN-E2SM-RC-v7.0} messages, supporting timely user-state reporting and frame delivery. xApps use E2 to ingest KPIs and configuration (channel state, throughput, utilization, energy) and to issue control commands. For completeness, O1 may also carry management directives related to O-CU-UP handling of generated frames.

We implement a RAN Resource Awareness (RRA) service model to report RAN KPIs and user channel state, and an RC service model to issue control commands (e.g., RB scheduling, transmit-power adjustments). The RRA feeds KPI/channel data to the XR xApp via E2, while RC conveys control actions to RAN nodes, forming the basis for real-time optimization in ImViD delivery.

\subsection{E2 Message Exchanges}

The E2AP protocol specifies interactions between the Near-RT RIC and O-CU/O-DU across four service types: Report, Insert, Control, and Policy. Service-model payloads are embedded within E2AP messages. We focus on Report and Control.

\begin{figure}[!htb]
\centering
\includegraphics[width=1\linewidth]{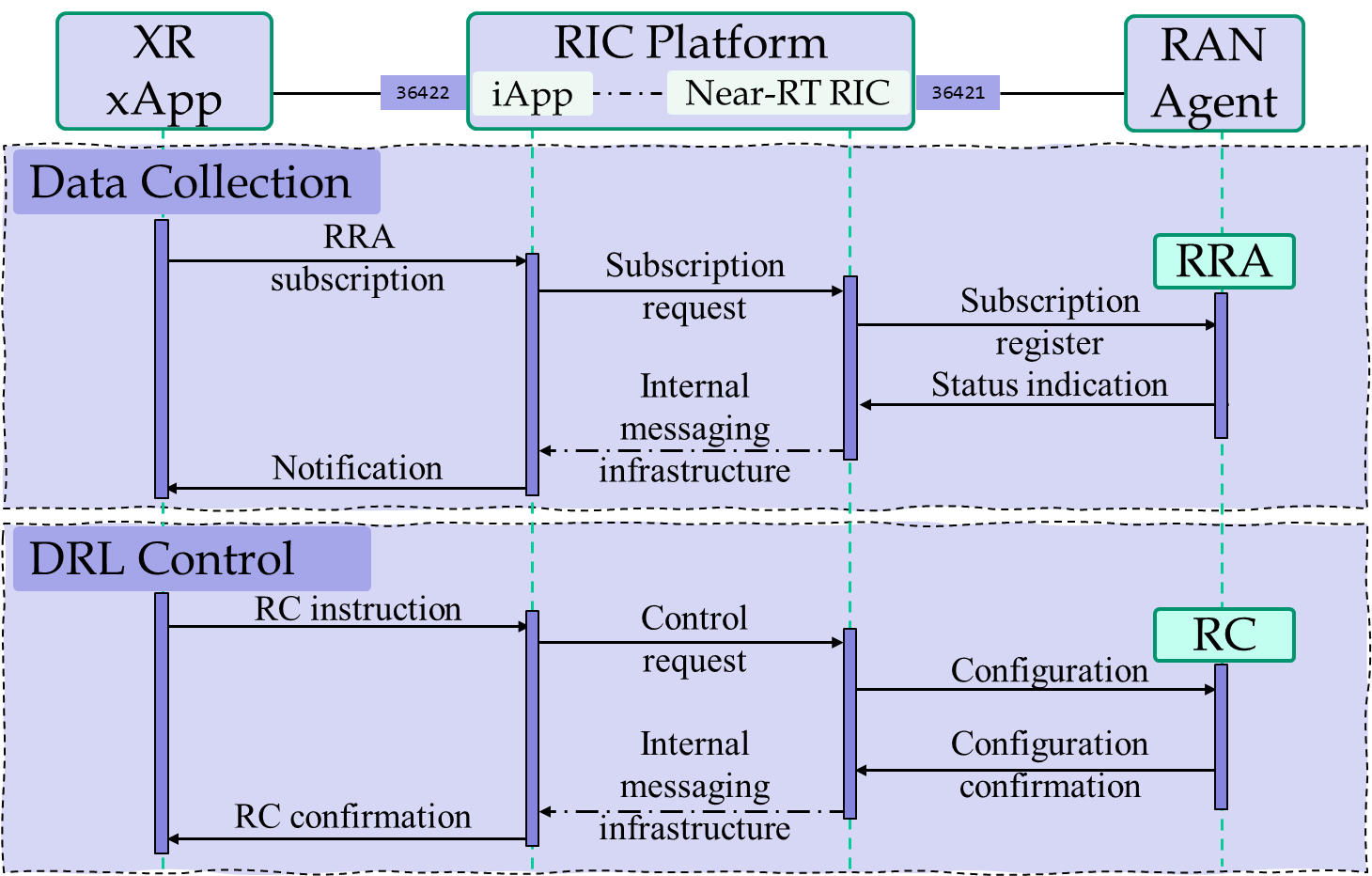}
\caption{Report and control message interaction process.}
\label{fig:e2_report_and_control}
\end{figure}
The Report service lets E2 nodes periodically report status, measurements, and statistics to the Near-RT RIC, supplying data for xApp decisions. Its interaction flow is shown in Fig.~\ref{fig:e2_report_and_control}.
After setup, the xApp subscribes the RRA service model on an E2 termination via the Near-RT RIC and receives a confirmation. The subscription defines periodic collection; when triggered, the E2 node packs RAN data and reports it to the RIC, which forwards it to the XR xApp for optimization.
The RC service enables the Near-RT RIC to issue control instructions to E2 nodes for radio-parameter adjustment, realizing closed-loop optimization. Its interaction flow is shown in Fig.~\ref{fig:e2_report_and_control}.
The XR xApp issues control requests based on reported data or internal logic. The E2 termination validates the request and distributes configuration to the target E2 node (e.g., RB and power allocation). On success, an acknowledgement is returned; if invalid or execution fails, a failure message is sent.
The online data collection and DRL-based control loops are implemented within an XR xApp running on the Near-RT RIC (FlexRIC). To facilitate interaction between the Python-based logic and the C/C++ components of the RAN within FlexRIC, a SWIG-produced Python wrapper is employed. This wrapper exposes the necessary application programming interfaces (APIs) in Python, which interact with the RAN through the standard O-RAN E2 procedures, specifically E2AP messages encapsulating E2SM payloads.

\section{Dynamic Resource Allocation by an xApp for Volumetric Video}

We consider a Time Division Duplexing (TDD) immersive system in which a single 5G gNB serves multiple HMD-equipped users. Prior to playback, users upload multi-view images to refresh the 3D scene model. During playback, each HMD periodically reports pose/position to an XR xApp, which requests novel-view rendering from the O-Cloud and schedules download delivery of the generated frames. Rendering and delivery operate in a closed loop until the session terminates. Consequently, upload traffic consists of (i) image sets for reconstruction and (ii) compact pose updates for playback, whereas download traffic carries latency-critical rendered frames. 
Building upon the O-RAN framework outlined in \cite{Li2025Rethinking}, we propose a playback system that concurrently orchestrates video generation, radio resource scheduling, and computational resources allocation in a Near-RT manner. The key objective is to optimize the trade-off between per-frame latency and rendering resolution in order to enhance the immersive QoE.

\subsection{Computation Model}

Both 3D scene reconstruction and novel-view generation are computationally intensive. Reconstruction starts from sparse structure-from-motion points and optimizes each 3D Gaussian's attributes to improve fidelity \cite{Xu2024Representing}. During playback, the xApp, given the user's instantaneous pose, computes the 2D covariance for each splatted 3D Gaussian \cite{Yifan2019Differentiable}, then renders pixels by aggregating Gaussian opacity/color \cite{Kopanas2021Point-Based,Kerbl20233D,Kopanas2022Neural}. Because pixel rendering is highly parallel, we treat each pixel as an independent task executed on the shared O-Cloud compute pool. Rendering time therefore scales with the number of rendered pixels (resolution) and the per-pixel workload. To accelerate rendering, \cite{Kerbl20233D} prunes low-density Gaussians and ignores distant Gaussians for a given viewpoint with negligible quality loss.

Let $\mathbf{v}\in\{0,1\}^{V}$ indicate rendered pixels in a frame with $V$ pixels, where $v_i=1$ denotes that pixel $i$ is rendered. Define $V'=\sum_{i=1}^{V}v_i$ and the content-hit ratio $\phi_u=V'/V\in[0,1]$ for user $u$. If $\Omega$ denotes the per-pixel compute (in cycles) and $f_u$ the processing frequency allocated to user $u$, the per-frame compute time is
\begin{align}
  T^{C}_{u} = \frac{V'\,\Omega}{f_{u}}.
\end{align}
Because pose updates determine the 2D covariances, state data must be refreshed in real time throughout rendering.

\subsection{Communication Model}

According to the Shannon formula, the instantaneous download rate on the gNB-user-$u$ link at slot $t$ is
\begin{equation} 
  R^{D}_{u}(t)= b^{D}_{u}(t) \log _{2} \left( {1+\frac {p^{D}_{u}(t)g_{u}(t)}{ \sigma ^{2}}} \right), 
\end{equation}
where 
$b^{D}_{u}(t)$ is the download bandwidth allocated to user $u$ at time $t$,
$p^{D}_{u}$ is the available transmission power for user $u$ at the gNB.
The signal-to-interference-plus-noise ratio (SINR) represents the channel state with $\frac {p^{D}_{u}(t)g_{u}(t)}{ \sigma ^{2}}$, where $\sigma ^{2}$ denotes the noise power.

Similarly, the maximum achievable upload transmission rate for user $u$ is 
\begin{equation} 
  R^{U}_{u}(t)= b^{U}_{u}(t) \log _{2}\left ({1+\frac {p^{U}_{u}(t) g_{u}(t)}{ \sigma ^{2}}}\right), 
\end{equation}
where 
$b^{U}_{u}(t)$ is the upload bandwidth allocated to user $u$ at time $t$, 
$p^{U}_{u}(t)$ is the upload power of user $u$ at time $t$. 
Therefore, the upload SINR is $\frac {p^{U}_{u}(t)g_{u}(t)}{ \sigma ^{2}}$ representing the channel state, $\sigma ^{2}$ denotes the noise power.

\subsection{Per-Frame Latency Decomposition}

Let $T_u$ be the end-to-end per-frame latency for user $u$ from pose upload to frame reception:
\begin{align}
T_{u} = T_{u}^{U} + T_{u}^{C} + T_{u}^{D},
\end{align}
where $T_{u}^{U}$, $T_{u}^{C}$ and $T_{u}^{D}$ are the upload transmission, computation, and download transmission delays, respectively.
Upload payloads are small status updates (e.g., head pose and position) and are not further compressible; therefore we focus on trading off computation and download volume by controlling the content-hit ratio $\phi_{u}$ (the fraction of pixels rendered).
Let $A_{u}^{U}$ be the VR status size, $A_{u}^{C}$ the computation required to render all pixels, and $A_{u}^{D}$ the download size for all pixels (no compression \cite{11311355}). Then
\begin{align}
T_{u}^{U} = \frac{A_{u}^{U}}{R_{u}^{U}}, 
T_{u}^{C} = \frac{A_{u}^{C}\,\phi_{u}}{f_{u}}, 
T_{u}^{D} = \frac{A_{u}^{D}\,\phi_{u}}{R_{u}^{D}},
\end{align}
where $R_{u}^{U}$ and $R_{u}^{D}$ are the upload and download rates and $f_{u}$ is the available processing rate. The factor $\phi_{u}\in[0,1]$ scales the computation and download volume.

\subsection{QoE Model and Optimization Problem}

User QoE depends on both latency and visual coverage. Following the Weber-Fechner law \cite{Lubashevsky2019Psychophysical,Reichl2010Logarithmic,Reichl2013Logarithmic}, we adopt
\begin{align}
  \mathrm{QoE}_u(t)=\left(1-\frac{T_u(t)}{T_{th}}\right)\ln\bigl(1+\phi_u(t)\bigr), \label{Formula:QoE}
\end{align}
where $T_{th}$ is the delay threshold, $T_u(t)$ is user $u$'s per-frame delay at slot $t$, and $\phi_u(t)\in[0,1]$ is the content-hit ratio.
We optimize the content-hit ratio and resource allocation over the following decision variables (indexed by slot $t$):
\begin{itemize}
  \item bandwidth allocations $\mathbf{B} = \{\mathbf{B}^{U},\mathbf{B}^{D}\}$ with $\mathbf{B}^{U}=\{b_u^{U}(t)\}_{u\in\mathbb{U}}$ and $\mathbf{B}^{D}=\{b_u^{D}(t)\}_{u\in\mathbb{U}}$;
  \item computation allocations $\mathbf{F}=\{f_u(t)\}_{u\in\mathbb{U}}$;
  \item download power allocations $\mathbf{P}=\{p_u^{D}(t)\}_{u\in\mathbb{U}}$;
  \item content-hit ratios $\Phi=\{\phi_u(t)\}_{u\in\mathbb{U}}$.
\end{itemize}
The long-term system objective maximizes the time-average expected sum-user QoE:
\begin{align} 
  \max_{ \textbf{B},\textbf{F},\textbf{P}, \Phi } & \lim_{T \rightarrow \infty } \frac{1}{T} \sum^{T-1}_{t=0} \mathbb{E} \left\{\mathrm{QoE}_{u}(t)\right\} \label{formula:optimizeQuestion} \\
  {\mathrm{ s.t.}} & \sum_{u \in \mathbb{U} } b_{u}^{D}(t) \leq b_{\mathrm {max}}, \tag{\ref{formula:optimizeQuestion}.1} \\
  & \sum_{u \in \mathbb{U} } b_{u}^{U}(t) \leq b_{\mathrm {max}}, \tag{\ref{formula:optimizeQuestion}.2} \\
  & \sum_{u \in \mathbb{U} } f_{u}(t) \leq f_{\mathrm {max}}, \tag{\ref{formula:optimizeQuestion}.3} \\
  & \sum_{u \in \mathbb{U}} p^{D}_{u}(t) \leq p_{\mathrm {max}}, \tag{\ref{formula:optimizeQuestion}.4}\\
  & 0 \leq \phi_{u} \leq 1, \forall u. \tag{\ref{formula:optimizeQuestion}.5}
\end{align}
Here $b_{\max}$, $f_{\max}$ and $p_{\max}$ denote the total download (and upload) spectral bandwidth budget, total computation budget and total download transmit power, respectively. The upload transmit power of each user is assumed fixed and treated as part of the user state when accessing the gNB.


\subsection{Reinforcement Learning-Based Solution}

The optimization involves multiple coupled variables: upload/download bandwidth, download power, the O-Cloud's compute allocation, and per-user content-hit ratio, forming a nonconvex, stochastic NLP that is NP-hard to solve optimally with classical methods.
To tackle this, we propose ImVol-DRL, a Deep Reinforcement Learning agent that learns resource and hit-ratio control policies in a dynamic environment. DRL naturally handles high-dimensional state/action spaces and adapts to time-varying channels and demands. ImVol-DRL aims to maximize long-term average user immersive experience while promoting fairness across users.

The state space, action space, and reward are defined under the RL framework as follows:

\textbf{State space}: The state space includes user channel gain, user uploaded data size, video rendering computation, video downstream data size, and user upload power. 
The state $\boldsymbol{S}(t)$ at each time slot $t$ can be defined as 
\begin{align}
  \boldsymbol{S}(t)=\{ g_{u}(t), A_{u}^{U}, A_{u}^{C}, A_{u}^{D}, p^{U}_{u}(t)\}, u \in \mathcal{U},
\end{align}
with the information acquired from the user HMD and RRA service models. 

\textbf{Action space}: The action space includes upload bandwidth allocation, computing resource allocation, downstream bandwidth allocation, download power allocation, and content-hit ratio.
The action vector of the whole system at the $t$ time slot can be formulated as
\begin{align}
  \boldsymbol{A}(t)=\{ b_{u}^{U}(t), f_{u}(t), b_{u}^{D}(t), p^{D}_{u}(t), \phi_{u}(t) \}, u \in \mathcal{U}.
\end{align}
We use $\mathcal{A}=\{ \textbf{B}^{U}, \textbf{F}, \textbf{B}^{D}, \textbf{P}, \Phi \}$ to denote the whole solution set. 

\textbf{Reward}: The DRL algorithm is designed to maximize the long-term average immersive experience of users while ensuring fairness among users.
The reward function of the whole system at time slot $t$ can be defined as
\begin{align}
  \begin{split}
    \boldsymbol{R}(t) & =\mathcal{R}(\boldsymbol{S}(t),\boldsymbol{A}(t)), \\
    & =\sum_{u=1}^{U}\text{QoE}_{u}(t)-\beta_{1} CoV(\text{QoE}(t)), \label{formula:rewardfunction}
  \end{split}
\end{align}
where $\mathrm{CoV}(\text{QoE}(t))$ is the coefficient of variation across users (standard deviation over mean) and $\beta_{1}$ tunes the fairness-efficiency tradeoff. Lower $\mathrm{CoV}$ implies more balanced user experiences.

\section{Simulations}

\subsection{Simulation Setting}
We evaluate the proposed resource orchestration using a Python-based simulator for multiuser immersive volumetric video. The learning-based SAC and DDPG are compared against three non-learning baselines (PF-AO, AVG and Cloud-AVG). We report both steady-state performance and reward dynamics during training.

In the simulations, the number of users varied from 2 to 16. Users cycled through 480P/720P/1080P/2K resolutions, with corresponding computation and communication demands of 10/20/30/40, respectively. Channel gains varied per slot in the range [0.5, 2.0]; noise power was set to 1. The upload transmit power was randomized to approximately 1 mW. The RAN had a 40 MHz bandwidth, a total download power of 10 W, and the O-Cloud compute budget was 10. The network latency from the RAN to the cloud was approximately 5 ms, which increased per-frame delay for cloud rendering. Experiments ran with Python 3.11 on an NVIDIA GeForce RTX 4090 GPU and Intel Core i9-14900K CPU.

\subsection{Benchmarks and Details}
We evaluate five benchmarks:
\begin{itemize}
\item \textbf{SAC:} Offline training is conducted over 200 episodes, each consisting of 20 steps. Each environment update triggers 10 gradient steps on both the actor and critic networks.
\item \textbf{DDPG \cite{Yu2024Attention-Based}:} The training schedule is identical to that of SAC to ensure a fair comparison.
\item \textbf{PF-AO \cite{Jalali2000Data}:} The Proportional Fairness-Alternative Optimization (PF-AO) algorithm is employed for resource allocation, aiming to optimize content-hit ratios.
\item \textbf{AVG:} A uniform allocation strategy is implemented for both communication and computation resources, with rendering performed at the O-Cloud.
\item \textbf{Cloud-AVG:} A similar uniform allocation strategy is applied, but with rendering conducted in the cloud, which introduces additional backhaul latency.
\end{itemize}
Unless otherwise noted, learning setups share the following defaults: replay buffer size 10{,}000, batch size 128, Adam optimizer with learning rate $3{\times}10^{-4}$, discount $\gamma{=}0.99$, soft target update $\tau{=}0.005$, and gradient clipping with max-norm $1.0$. Each update performs 10 internal gradient iterations.

\subsubsection{SAC setting}

For SAC, we use a two-hidden-layer MLP actor (256-unit ReLU layers) and twin Q-critics with target networks and $\tanh$-squashed actions. The standard SAC objectives are used with the minimum over two critics and entropy regularization.
Concretely, the losses implemented are equivalent to
\begin{align}
& y = r + \gamma\Big(\min_{i=1,2} Q'_i(s',a') - \alpha \log\pi_\theta(a' \mid s')\Big), \label{y_func_sac} \\
& L_{critic} = \mathbb{E}\big[(Q(s,a) - y)^2\big], \label{L_critic_sac} \\
& L_{actor} = \alpha \log\pi_\theta( \tilde{a} \mid s) - \min_{i=1,2} Q_i(s,\tilde{a}), \label{L_actor_sac} \\
& L_{alpha} = - \alpha \mathbb{E}\big[\log\pi_\theta(a \mid s) + \mathcal{H}_{target}\big], \label{L_alpha_sac}
\end{align}
where $\alpha=0.2$ is the entropy temperature, $r$ is reward, $s$ is the current state, $s'$ is the next state, $a$ is the current action, $a'$ is the next action, $\tilde{a}$ is the noise random variable, actor parameters $\theta$, $Q'$ is the target Q critic, $Q$ is the current Q critic, $\mathcal{H}_{target}$ is the target entropy setting to $-\lvert \mathcal{A}\rvert$ (negative action dimensionality), $\pi$ is policy function.
Numerical safeguards include log-std clamping and a small $\epsilon$ in $\tanh$-related log-probabilities to avoid $\log(0)$.

It is important to highlight that power, bandwidth, computational resources, and the content-hit ratio are continuous variables, subject to upper bounds. In cases where the SAC outputs infeasible actions, these are projected and rescaled into the feasible set, with a penalty applied. Specifically, for power, bandwidth, and computation, which are constrained by hard sum budgets, the actions are passed through a $\tanh$ function, mapped to the interval $(0,1)$, and then normalized across users to derive nonnegative allocation ratios that sum to unity. These ratios are subsequently scaled by the respective total budgets. The content-hit ratio is treated as an independent, continuous, and bounded action, which is passed through a $\tanh$ function, mapped to $[0,1]$, and clipped within predefined limits to ensure it remains within a reasonable range. To promote fairness while respecting hard constraints, we introduce an additional penalty on the dispersion of resources across users, quantified by the coefficient of variation, which is weighted by a fairness factor $\beta_1$.

Additionally, we provide the pseudo-code for Algorithm \ref{alg:sac_training}, which outlines the main SAC agent loop. This loop encompasses the processes of state observation, action selection, environment interaction, experience storage, and network updates.
\begin{algorithm}[t]
\caption{Soft Actor-Critic (SAC) training procedure.}
\label{alg:sac_training}
\begin{small}
\begin{algorithmic}[1]
  \STATE \textbf{Input:} learning rates $\eta_\pi, \eta_Q, \eta_\alpha$, discount factor $\gamma$, target smoothing coefficient $\tau$, target entropy $\mathcal{H}_{\text{target}}$, batch size $B$.
  \STATE \textbf{Initialize:} actor parameters $\theta$, critic parameters $\phi_1, \phi_2$, target critic parameters $\bar\phi_1 \leftarrow \phi_1$, $\bar\phi_2 \leftarrow \phi_2$, entropy temperature $\alpha$ (learnable), replay buffer $\mathcal{D}$.
  \FOR{each environment step (episode loop outside)}
    \STATE Observe state $s$.
    \STATE Sample action $a \sim \pi_\theta(\cdot|s)$.
    \STATE Execute action $a$, observe reward $r$ and next state $s'$.
    \STATE Store transition $(s,a,r,s')$ in replay buffer $\mathcal{D}$.
    \IF{$|\mathcal{D}| \geq \text{batch\_size}$}
      \STATE Sample minibatch $(s,a,r,s')$ from $\mathcal{D}$.
      \STATE Sample target actions $a' \sim \pi_\theta(\cdot|s')$.
      \STATE Calculate policy $\pi_\theta$'s loss using KL divergence in \eqref{y_func_sac}.
      \STATE Calculate target Q with \eqref{y_func_sac}.
      \STATE Update critics by minimizing \eqref{L_critic_sac}.
      \STATE Update actor by minimizing \eqref{L_actor_sac}.
      \IF{$\alpha$ is learnable}
        \STATE Update $\alpha$ by minimizing \eqref{L_alpha_sac}.
      \ENDIF
      \STATE Soft-update target critics: $\bar\phi_i = \tau \phi_i + (1-\tau)\bar\phi_i, \forall i\in\{1,2\}$.
    \ENDIF
  \ENDFOR
\end{algorithmic}
\end{small}
\end{algorithm}

\subsubsection{DDPG setting}

For DDPG \cite{Yu2024Attention-Based}, the actor has two hidden layers (400, 300 units) and outputs actions of relu activation and tanh scaling, which are then mapped to $[0,1]$; the critic concatenates state and action to produce a scalar $Q$. Exploration uses Gaussian noise with standard deviation $\approx 0.1$. Target networks use soft updates with the same $(\gamma,\tau)$ as above. Losses follow the standard MSE target for the critic and a deterministic policy gradient that maximizes $Q$, i.e.
\begin{align}
& L_{critic} = \mathbb{E}\big[(Q(s,a) - Q'(s',a'))^2\big], \\
& L_{actor} = -Q(s,\tilde{a}).
\end{align}
The reward function, training procedure, and update steps of DDPG are similar to those of SAC, while the main difference lies in the fact that DDPG does not have a target Q Critic. Meanwhile, DDPG handles the action space in the same way as SAC, using a linear mapping to project actions into the feasible region so as to satisfy both the total resource budget constraints and the bounds of independent variables. 
The DDPG pseudo-code is similar to SAC’s: remove the target Q-critic steps from the SAC training pseudo-code and replace equations in lines 13 and 14 of Algorithm \ref{alg:sac_training} with equations (15) and (16), respectively; further details are omitted for brevity.

\subsubsection{PF-AO setting}

Proportional Fairness (PF) is widely used for resource allocation to balance throughput and user fairness by maximizing the sum of logarithmic user data rates \cite{Jalali2000Data}. We compute per-user PF coefficients from the ratio between instantaneous demand and historical average throughput, and allocate resources proportionally to capture both short-term demand and long-term fairness.

\begin{figure}
\centering
\begin{align}
  \mathrm{QoE}_u =& \left( 1-\frac{ {T_{u}^{U} + \frac{(A_{u}^{C}}{f_{u}}  + \frac{A_{u}^{D}}{R_{u}^{D}}) \phi_{u}} }{T_{th}} \right) \ln\bigl(1+\phi_u\bigr). \label{eq:diff_QoE} \\
 \frac{ \partial \mathrm{QoE}_u }{ \partial \phi_u } =& \left(1 - \frac{ T_{u}^{U} + (A_{u}^{C}/f_{u} + A_{u}^{D}/R_{u}^{D}) \phi_{u} }{T_{th}} \right) \frac{1}{1+\phi_u} \notag \\ & -\frac{(A_{u}^{C}/f_{u} + A_{u}^{D}/R_{u}^{D})}{T_{th}} \ln(1+\phi_u), \\
 = & \left(1 - \frac{ T_{u}^{U} - (A_{u}^{C}/f_{u} + A_{u}^{D}/R_{u}^{D}) }{T_{th}} \right) \frac{1}{1+\phi_u} \notag \\ & - \frac{ (A_{u}^{C}/f_{u} + A_{u}^{D}/R_{u}^{D})}{T_{th}} \notag \\ & -\frac{(A_{u}^{C}/f_{u} + A_{u}^{D}/R_{u}^{D})}{T_{th}} \ln(1+\phi_u). \label{eq:diff_QoE2} \\
 \frac{ \partial \mathrm{QoE}_u^2 }{ \partial^2 \phi_u } =& \left(1 - \frac{ T_{u}^{U} + (A_{u}^{C}/f_{u} + A_{u}^{D}/R_{u}^{D}) }{T_{th}} \right) \frac{-1}{(1+\phi_u)^2} \notag \\ & -\frac{(A_{u}^{C}/f_{u} + A_{u}^{D}/R_{u}^{D})}{T_{th} (1+\phi_u)} < 0. \label{eq:diff_QoE2_less} \\
\frac{ \partial \mathrm{QoE}_u }{ \partial \phi_u } & {\Big|} _{\phi_u = 0} = \left(1 - \frac{ T_{u}^{U} }{T_{th}} \right) > 0. \label{eq:diff_QoE_greater}
\end{align}
\end{figure}
The QoE function in \eqref{eq:diff_QoE} admits a unique maximizer over $\phi_u\in[0,1]$, which follows from the derivative analysis in \eqref{eq:diff_QoE}-\eqref{eq:diff_QoE2} \cite{Wen2025Training}. With $T_u\leq T_{th}$, QoE remains nonnegative, and $\partial^2\mathrm{QoE}_u/\partial\phi_u^2<0$ in \eqref{eq:diff_QoE2_less} together with \eqref{eq:diff_QoE_greater} implies a single optimum given by solving $\partial\mathrm{QoE}_u/\partial\phi_u=0$ (or the boundary when needed). We compute it numerically using CVX \cite{Zhang2025Enhancing}, and Fig.~\ref{fig:func_visual_latency_phi} visualizes the resulting QoE-hit-ratio tradeoff.

In our problem, rendering compute and download traffic scale with the content-hit ratio, so we propose a Proportional Fairness-Alternating Optimization (PF-AO) method in Algorithm~\ref{alg:pf_ao}. It alternates between PF-based allocation of upload bandwidth, O-Cloud compute, and download bandwidth, and CVX-based optimization of per-user content-hit ratios under the current allocation; the detailed PF coefficient derivation follows \cite{Jalali2000Data}.
\begin{algorithm}[t]
\caption{PF-AO decision making procedure.}
\label{alg:pf_ao}
\begin{small}
\begin{algorithmic}[1]
  \STATE Initialize download power allocation ratios.
  \STATE Compute PF coefficients for upload traffic, allocate upload bandwidth proportionally.
  \FOR{Alternating Optimization loop}
    \STATE Compute per-user rendering compute demand and download traffic demand.
    \STATE Compute PF coefficients for rendering compute and download traffic, allocate O-Cloud compute and download bandwidth proportionally.
    \STATE Solve the content-hit ratio by CVX to maximize the sum QoE across users under the current resource allocation.
  \ENDFOR
\end{algorithmic}
\end{small}
\end{algorithm}
Algorithm~\ref{alg:pf_ao} fixes equal download-power ratios, alternates the PF allocation and CVX-based hit-ratio optimization \cite{Wen2025Training}, and runs for 10 iterations in our simulations.

\subsection{Simulation Results and Analysis}



In this section, we investigate how latency relates to immersive QoE and the content-hit ratio, and examine the training dynamics of different algorithms through reward evolution and per-user latency variation. We also compare average immersive QoE, content-hit ratio, scheduling success rate, and average latency under different resolution settings, and evaluate these metrics in multi-user scenarios. While PF-AO attains a higher reward in the early stage of small-scale simulations, SAC steadily improves during training and reaches a comparable reward. Under multi-user, high-load conditions, SAC achieves lower average latency and better average QoE, and its inference time is much smaller than PF-AO in practical deployment, indicating stronger long-term decision refinement, better global resource optimization, and improved stability for Near-RT per-frame resource allocation.

\subsubsection{Latency and QoE on Content-Hit Ratio}

\begin{figure}[!htb]
    \centering
    \includegraphics[width=1\linewidth]{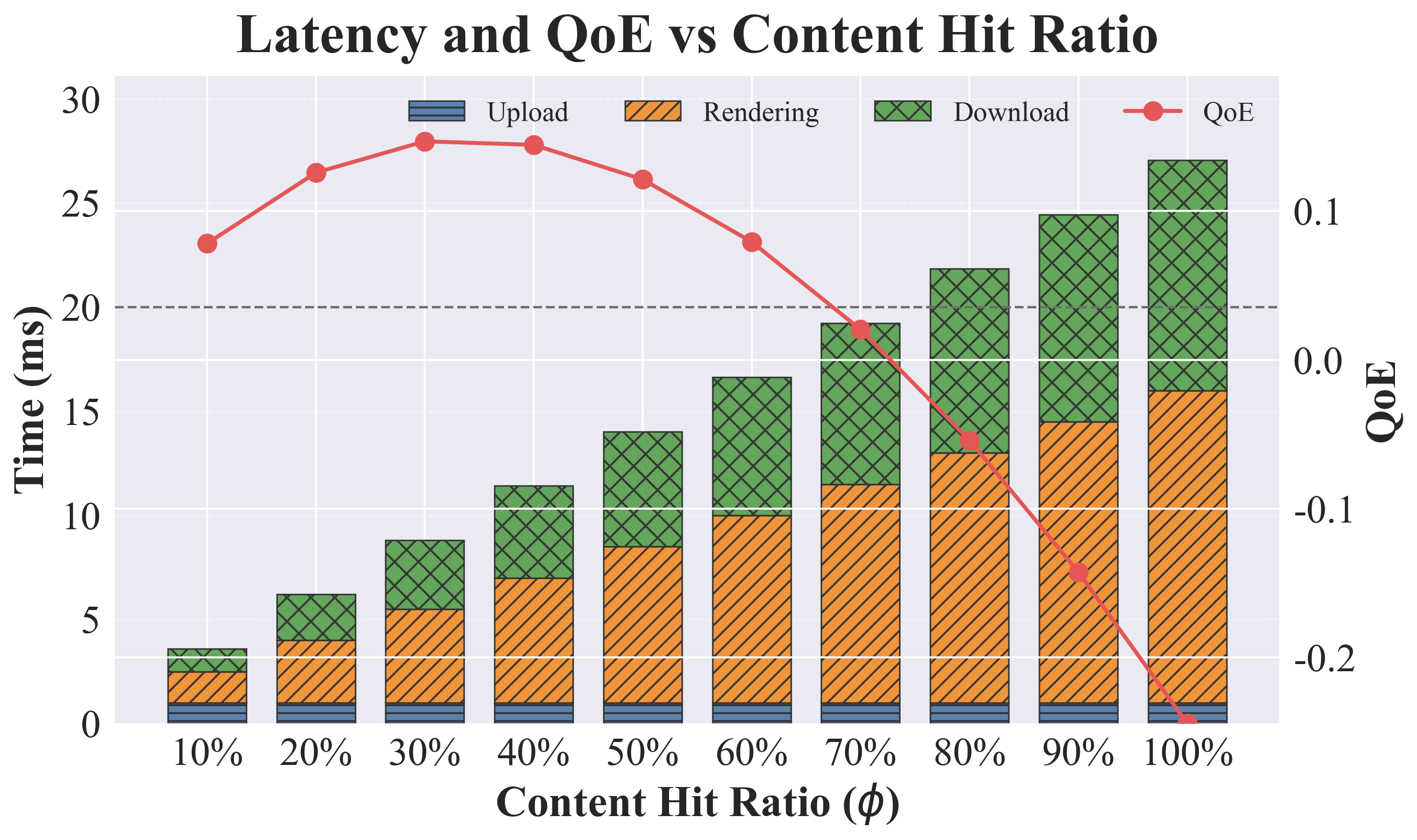}
  \caption{Latency and QoE versus content-hit ratio.}
  \label{fig:func_visual_latency_phi}
\end{figure}

To provide a more intuitive understanding, Fig.~\ref{fig:func_visual_latency_phi} visualizes the impact of the content-hit ratio on latency. 
Using the content-hit ratio as the horizontal axis, we plot a stacked bar chart of the end-to-end latency components. The results show that, as the content-hit ratio increases, both computation delay and download delay grow significantly, leading to a higher end-to-end latency. In our experimental setting, the upload delay is relatively small and independent of the content-hit ratio, so its contribution to the overall latency is minor.
Specifically, when the content-hit ratio increases from 0\% to 30\%, the visual-quality gain dominates and the user QoE improves. Beyond 30\%, however, the latency penalty grows rapidly and the marginal QoE benefit from higher rendered coverage cannot offset the increasing end-to-end delay, so QoE starts to decrease. When the content-hit ratio is between 30\% and 70\%, the end-to-end latency stays within the 20 ms threshold, so the immersive experience can be maintained. When the content-hit ratio exceeds 70\%, the end-to-end latency goes beyond the immersive threshold, i.e., the scheduling fails, and the QoE becomes negative.

In general, when the content-hit ratio falls below 50\%, the effective visual resolution becomes too low and the perceived visual experience degrades noticeably. A more suitable strategy is to adapt the nominal resolution configuration to improve immersion.

\subsubsection{Training Process}

\begin{figure}[!htb]
    \centering
    \includegraphics[width=1\linewidth]{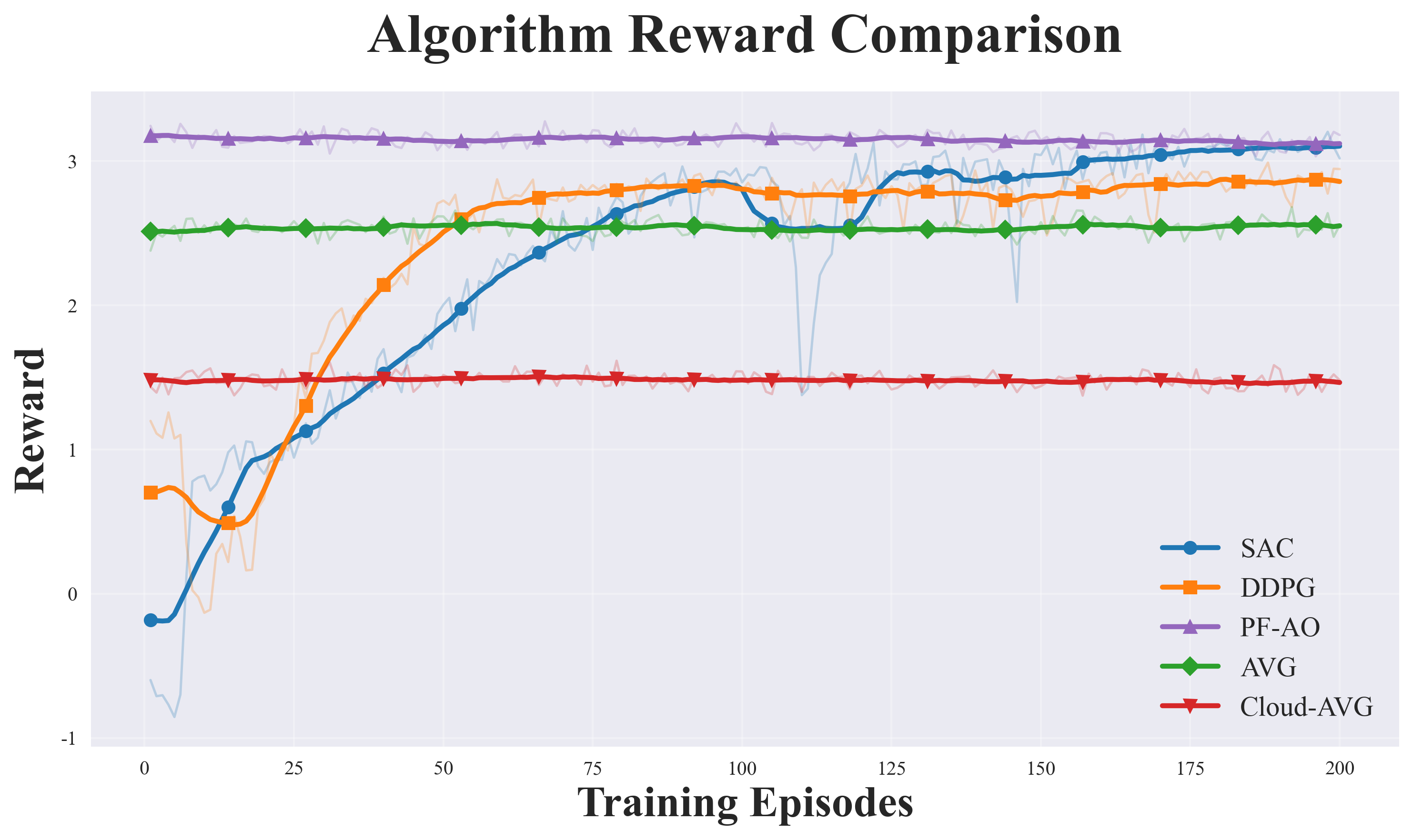}
  \caption{Reward evolution during training.}
  \label{fig:simulation_reward}
\end{figure}
Fig.~\ref{fig:simulation_reward} shows the reward evolution during training with eight users. Overall, rewards increase with iterations and stabilize as policies improve. 
Early fluctuations arise from unoptimized policies; experience replay and target networks in DDPG/SAC enhance sample efficiency and learning stability.
DDPG converges faster at the start but exhibits larger oscillations and attains a lower final reward, likely due to aggressive updates and noise sensitivity. 
SAC, by maximizing entropy, explores more broadly and achieves smoother learning with a higher final reward. 
In contrast, AVG and Cloud-AVG change little, indicating limited adaptability to dynamic environments. 
Compared with cloud rendering, the O-Cloud scheme delivers lower latency and higher final rewards, while cloud rendering incurs frame-transmission delays, and the O-Cloud's flexible resource allocation better serves real-time user demands.
Although PF-AO achieves a higher reward in the early stage, SAC gradually improves and reaches a comparable reward as training proceeds. This indicates that SAC can continuously learn more effective decisions for long-term user-experience optimization and exhibits stronger capability.

\begin{figure}[!htb]
  \centering
  \includegraphics[width=1\linewidth]{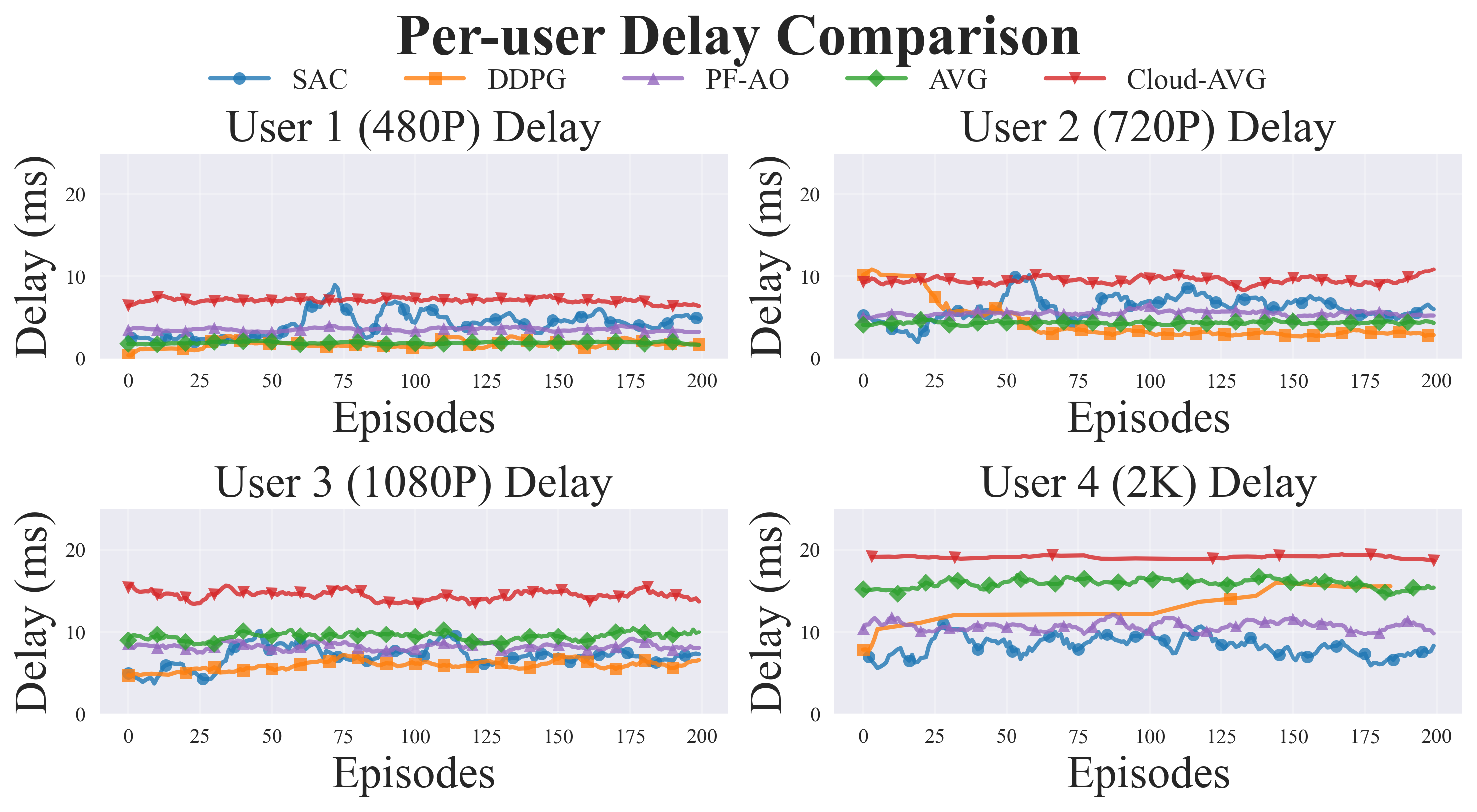}
  \caption{Per-user delay reduction during training.}
  \label{fig:simulation_per_user_delay_reduction}
\end{figure}
In a concise training-dynamics figure (Fig.~\ref{fig:simulation_per_user_delay_reduction}), we illustrate how per-user latency evolves over training episodes for SAC, DDPG and PF-AO. 
The figure shows that SAC actively explores candidate solutions in the early stage while maintaining good overall performance, which leads to larger latency fluctuations across users until around 125 episodes. 
For User 4 (2K), whose computation and communication demands are high and thus place heavy pressure on limited RAN resources, SAC gradually converges as training proceeds and achieves the lowest latency among the compared methods, while keeping the latency of other users from becoming noticeably higher than that under the baselines. Overall, these results highlight SAC’s training stability and its practical advantage in learning faster and improving user experience more effectively.

\subsubsection{Average Immersive QoE, Hit Ratio, and Latency Performance}

\begin{figure}[!htb]
    \centering
    \includegraphics[width=1\linewidth]{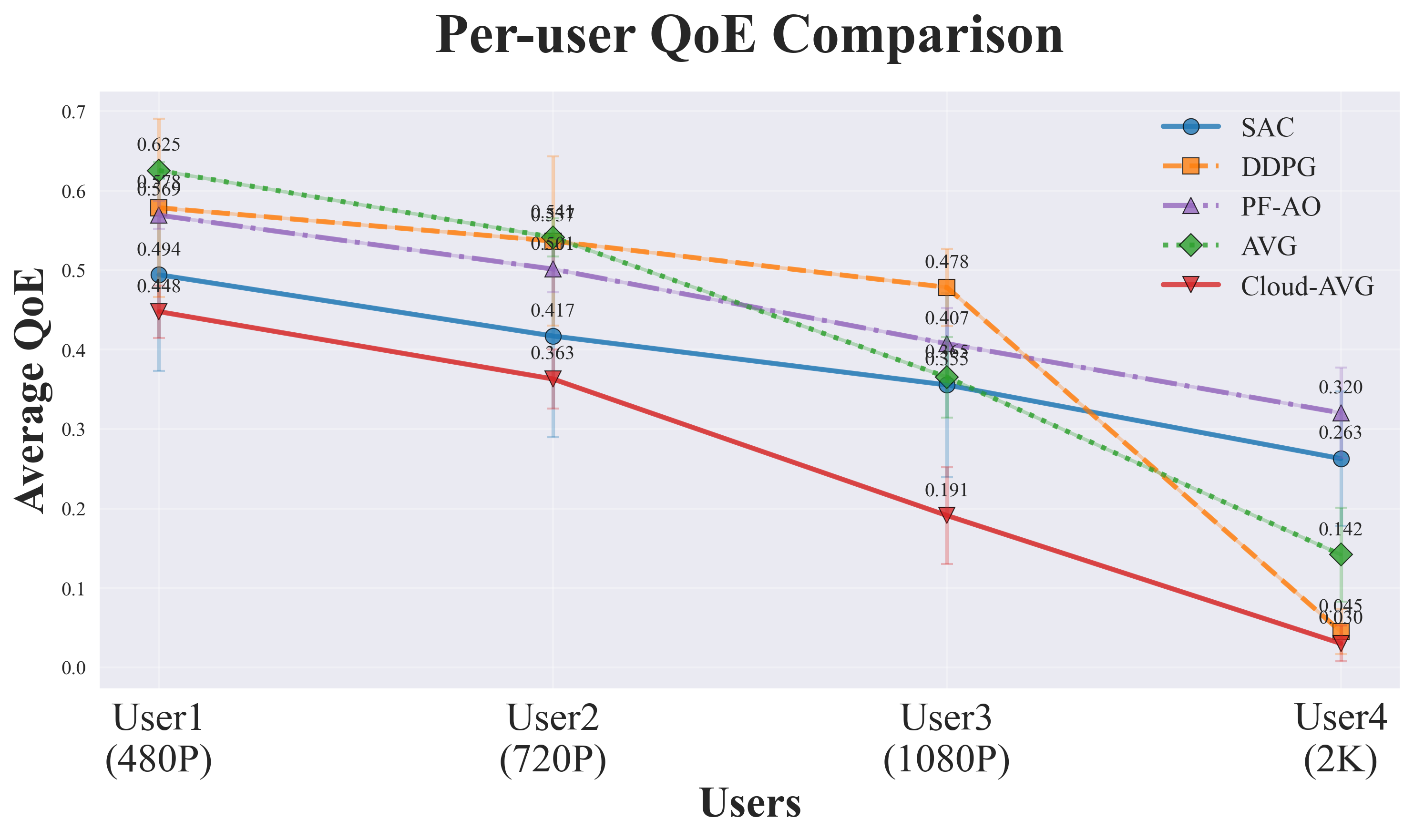}
  \caption{Per-user QoE comparison.}
\label{fig:simulation_qoe}
\end{figure}
Fig.~\ref{fig:simulation_qoe} compares the QoE of the first four users under different algorithms in an eight users simulation. 
QoE decreases as resolution rises (480p to 2K). 
At low resolutions, DDPG and AVG yield higher QoE than SAC, DDPG's replay favors frequent low-resolution successes, and AVG's equal resource allocation benefits low-demand users. For high-resolution users, SAC surpasses AVG by continuously exploring policies that raise overall QoE. Although SAC slightly reduces QoE for low-resolution users, it delivers higher overall QoE with smaller inter-user disparities and a more stable, robust allocation than DDPG and AVG.
SAC does not outperform PF-AO in terms of overall QoE under the eight-user setting. This is mainly because PF-AO optimizes resource allocaiton and the content-hit ratio using demand-driven allocation and an accurate convex optimization solver, which leads to a more effective resource allocation.
This also highly reveals a limitation of PF-AO. It does not adapt well to dynamic environments and it tends to keep allocating resources to every user even under high load. In the subsequent multi-user experiments, when the number of users increases to 16, the advantage of SAC under high load becomes more evident. 
Although SAC does not achieve a higher scheduling success rate than PF-AO, it attains a higher average QoE. Moreover, as user resource demands increase, the QoE gap between SAC and PF-AO becomes smaller. This indicates that SAC has strong global resource allocation capability and can better handle such high-dimensional resource allocation problems.

\begin{figure}[!htb]
    \centering
    \includegraphics[width=1\linewidth]{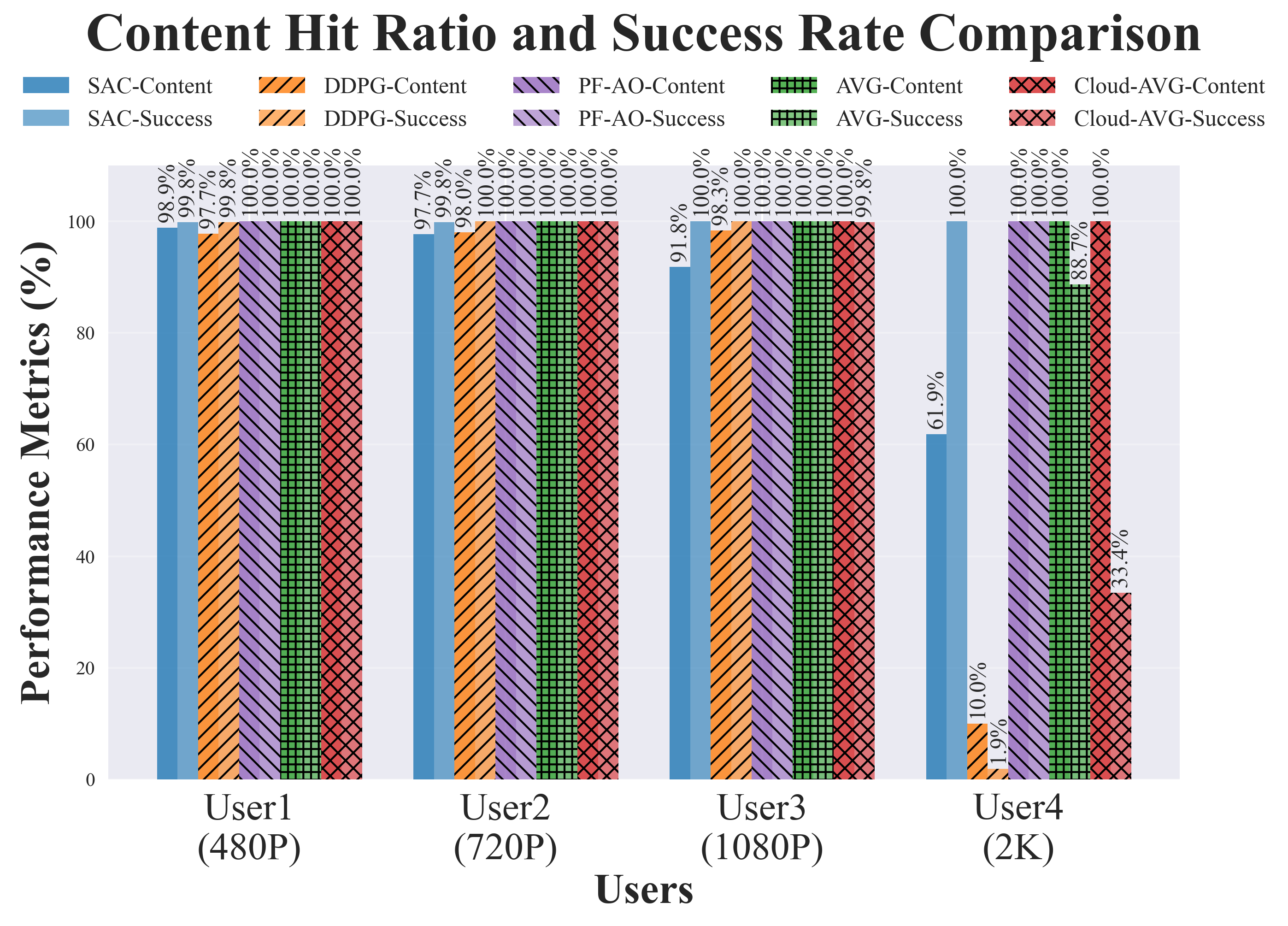}
  \caption{Content-hit ratio and successful scheduling rate.}
  \label{fig:simulation_hit_ratio}
\end{figure}
In Fig.~\ref{fig:simulation_hit_ratio}, as the resolution configuration selected by the user increases (from 480P to 2K), the content-hit ratio of the system shows a downward trend.
The SAC algorithm outperforms the average allocation algorithm in terms of content-hit ratio when users choose high resolution, indicating that the SAC algorithm can more intelligently adapt to user needs and improve the overall performance of the system.

\begin{figure}[!htb]
    \centering
    \includegraphics[width=1\linewidth]{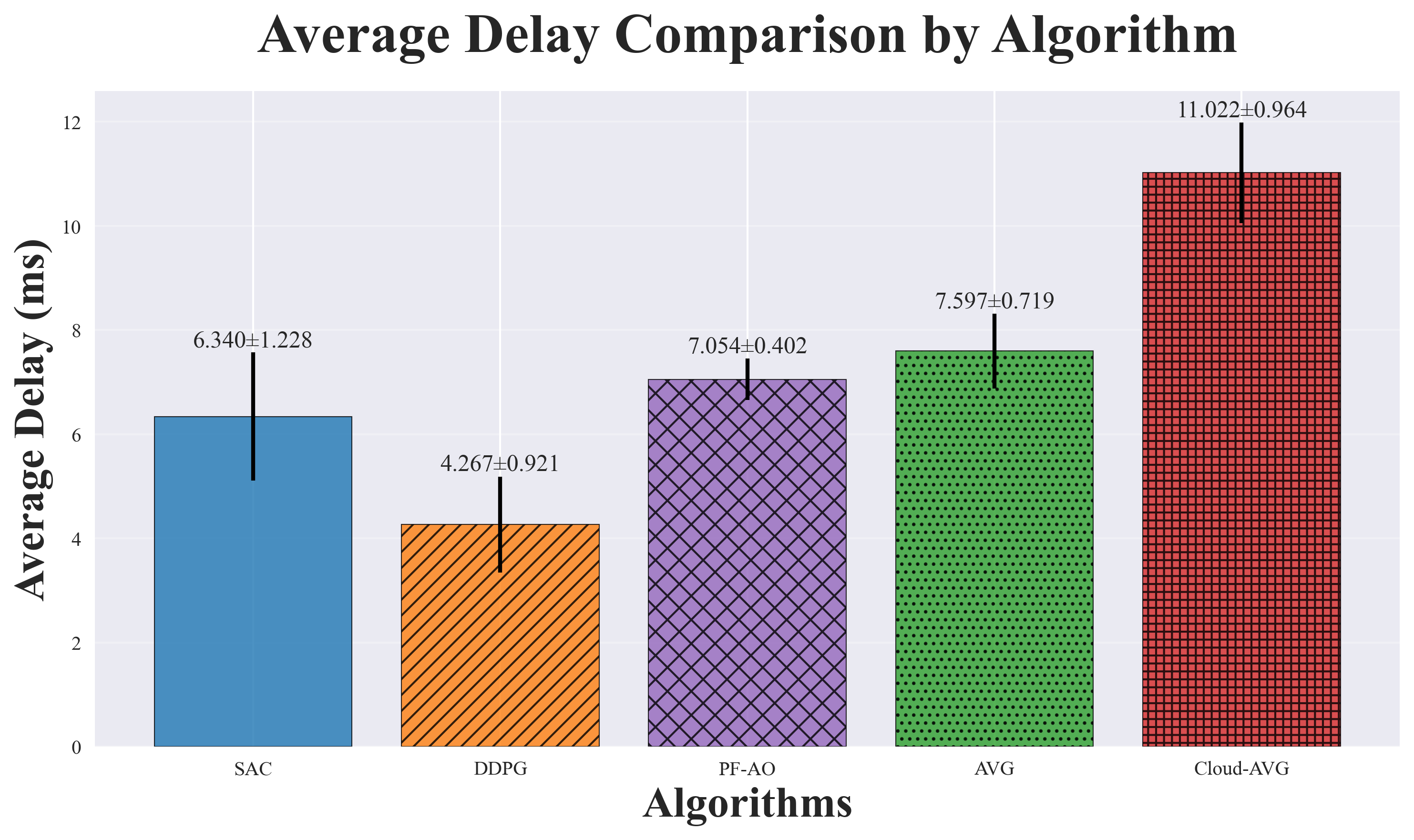}
  \caption{Average per-frame latency under different policies.}
  \label{fig:simulation_delay}
\end{figure}
In Fig.~\ref{fig:simulation_delay}, the average latency of the SAC algorithm is lower than that of the AVG allocation algorithm, $19\%$ lower than that of the O-Cloud-based average allocation algorithm, $73\%$ lower than that of the Cloud-AVG algorithm, and $11\%$ lower than that of the PF-AO algorithm.
It can be seen that the average latency of the DDPG algorithm is lower than that of the SAC algorithm. We can also see this reasonable explanation from Fig.~\ref{fig:simulation_qoe} and Fig.~\ref{fig:simulation_hit_ratio}, which is due to the prioritization of low resolution users in the DDPG algorithm, thereby reducing the overall average latency. Therefore, adopting the SAC algorithm can effectively control the average delay of the system while improving user QoE, and has robustness.

\subsubsection{Impact of User Number}

\begin{figure}[!htb]
  \centering
  \includegraphics[width=1\linewidth]{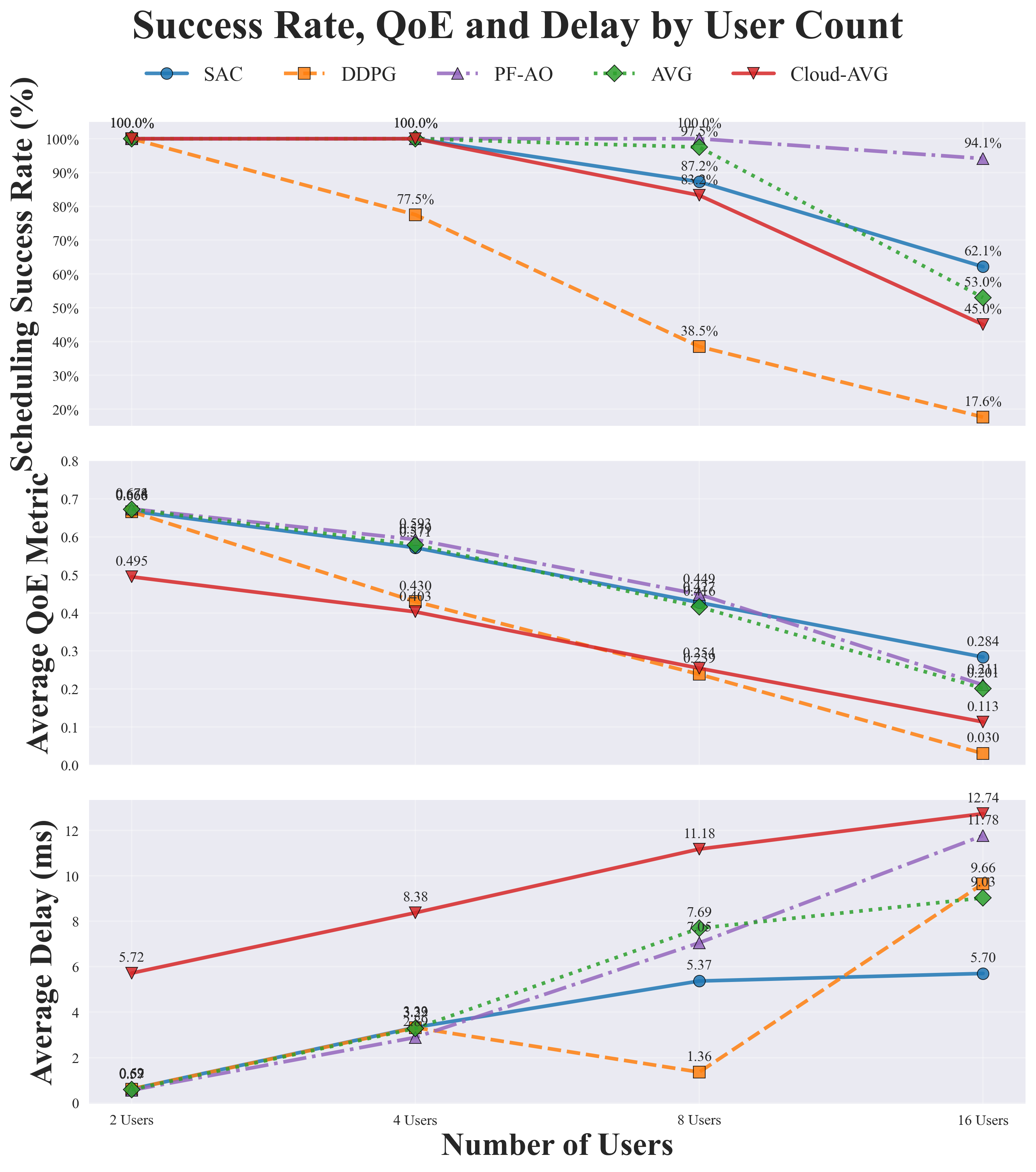}
  \caption{Success rate, QoE and Delay versus user number.}
  \label{fig:simulation_success_rate_qoe_delay}
\end{figure}

In Fig.~\ref{fig:simulation_success_rate_qoe_delay}, as the number of users increases, compute and communication resources become scarce, causing a decline in allocation success rates. 
SAC outperforms the other methods at larger scales and remain more stable across different user counts. 
As the user population grows, all methods see reduced success rates and QoE, but SAC stays more adaptive under heavier loads, especially with 16 users, and consistently outperforms DDPG and static baselines in QoE. This also suggests that the reward in Eq.~(\ref{formula:rewardfunction}) effectively guides the policy to improve overall QoE while promoting fairness among users.
The average latency increases for all methods as the number of users grows, while SAC shows a smaller increase, indicating more effective resource management under scarce resources. In contrast, DDPG increases more sharply, particularly at 8 and 16 users, and exhibits larger fluctuations, reflecting reduced stability under high load; the static allocation baseline yields the highest latency, further confirming its limited adaptability.
SAC can better cope with high-dimensional resource allocation under heavy load.

\section{Prototype System and Analysis}

\subsection{Prototype System Implementation}

\begin{figure}[!htb]
  \centering
  \includegraphics[width=1\linewidth]{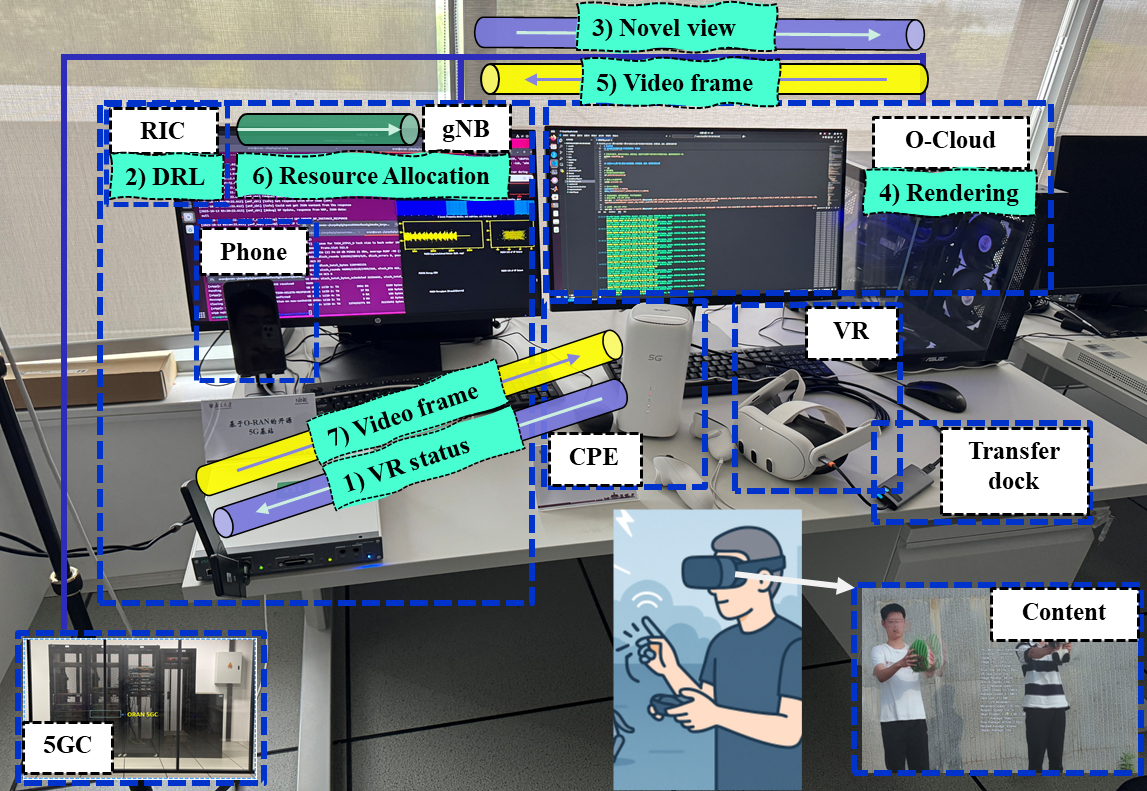}
  \caption{Deployment of the XR O-RAN communication system.}
  \label{fig:deploy_vr_oran}
\end{figure}

We built an open-source 5G testbed composed of an XR device, a Radio Unit (RU), a Distributed Unit and a Centralized Unit (DU/CU), a RIC, and a 5G Core (5GC). 
The system follows the O-RAN architecture standard with specifications in Tab.~\ref{tab:prototype_spec}. The UE supports dual-mode operation: a commercial off-the-shelf smartphone with a physical SIM card and an emulated OpenAirInterface (OAI) nrUE connection. 
The RU uses an Ettus USRP X300 with one UBX-160 daughterboard, which supports sub-6GHz frequency with two rubber rod antennas. The DU, CU and RIC are all deployed on x86-64 desktop servers running Ubuntu 22.04.5 LTS with the 6.8.0-48-lowlatency kernel, 32 GB of RAM, and an Intel i7-13700 CPU. 
The gNB software is based on OAI version \textit{2024.w02}, and the RIC implementation is based on FlexRIC at commit \textit{035fd2e8}. 
The core network uses the OAI 5G CN project at commit \textit{f1d9a95d}. All core network components run in Docker containers on rack-mounted servers in our data center (bottom left corner in  Fig.~\ref{fig:deploy_vr_oran}). 
The access network operates in TDD mode on the n78 band (3.3 GHz) with a 40 MHz channel and implements the full 5G protocol stack; it supports Internet access for UEs with peak download throughput up to 116 Mbps.
The O-Cloud platform is deployed on rack-mounted servers, and logs are displayed on the desktop computer in the top right corner of Fig.~\ref{fig:deploy_vr_oran}. The server CPU is an Intel(R) Xeon(R) Platinum 8336C CPU @ 2.30GHz, and the GPU is NVIDIA GeForce RTX 4090.
\begin{table}[!t]
\centering
\caption{Prototype software and hardware specifications.}
\label{tab:prototype_spec}
\begin{tabularx}{\linewidth}{p{2cm}X}
\toprule
  \textbf{Item} & \textbf{Specification} \\
\midrule
\multicolumn{2}{l}{\textbf{Software}} \\
OS & Ubuntu $22.04.5$ LTS, kernel $6.8.0$-$48$-lowlatency \\
gNB & OAI (OpenAirInterface) \textit{2024.w02} \\
RIC & FlexRIC, commit \textit{035fd2e8} \\
5GC & OAI 5G CN, commit \textit{f1d9a95d} \\
Containerization & Docker 28.0.1, build 068a01e \\
GCC & GCC version $10.5.0$ \\
Python & Python version $3.8.20$ \\
E2 Message & Indication, Control service \cite{ORAN-RICARCH-v7.0,ORAN-E2SM-KPM-v6.0,ORAN-E2SM-RC-v7.0} \\

\addlinespace
\multicolumn{2}{l}{\textbf{Hardware}} \\
RU & Ettus USRP X300 + UBX-160 daughterboard \\
Antennas & Two rubber rod antennas (sub-6 GHz) \\
gNB server & Intel i7-13700 CPU, RAM $32$ GB \\
O-Cloud server & Intel Xeon 8336C @ $2.30$ GHz, RAM $128$ GB, NVIDIA RTX 4090 \\
RAN & TDD, n78 ($3.3$ GHz), $40$ MHz channel \\
CPE & FiberHome 5G CPE LG6851F \\
VR & Meta Quest 3 \\
\bottomrule
\end{tabularx}
\end{table}

In our setup, as shown in Fig.~\ref{fig:deploy_vr_oran}, a Meta Quest 3 HMD is connected to a customer premises equipment (CPE) client via a dock and an RJ-45 Ethernet cable. The SMO is deployed on a rack mounted server (shown at the upper-right machine in figure), which also hosts the O-Cloud platform; this desktop provides the primary compute resources for cloud-side services. The gNB runs on the upper-left desktop in the picture. 
The CPE connects to our gNB over the 5G radio network. We provisioned a SIM card programmed with PLMN 00101, inserted it into the CPE, and locked the CPE's radio to the n78 frequency band. The CPE and the HMD are connected via the RJ-45 port on the dock; the forwarding and routing delay introduced by the dock and cable is negligible. The CPE handles the radio link to the gNB, uploads status and application data from the HMD to the gNB, and forwards download video frame traffic from the gNB to the HMD. Thus, the XR device attains full Internet connectivity through the deployed 5G network to stream the ImViD.

\subsection{Time Consumption Analysis}

\begin{figure}[!htb]
  \centering
  \subfigure[Per-frame time consumption of playback.\label{fig:deploy_vr_time_consumption}]{
    \includegraphics[width=0.95\linewidth]{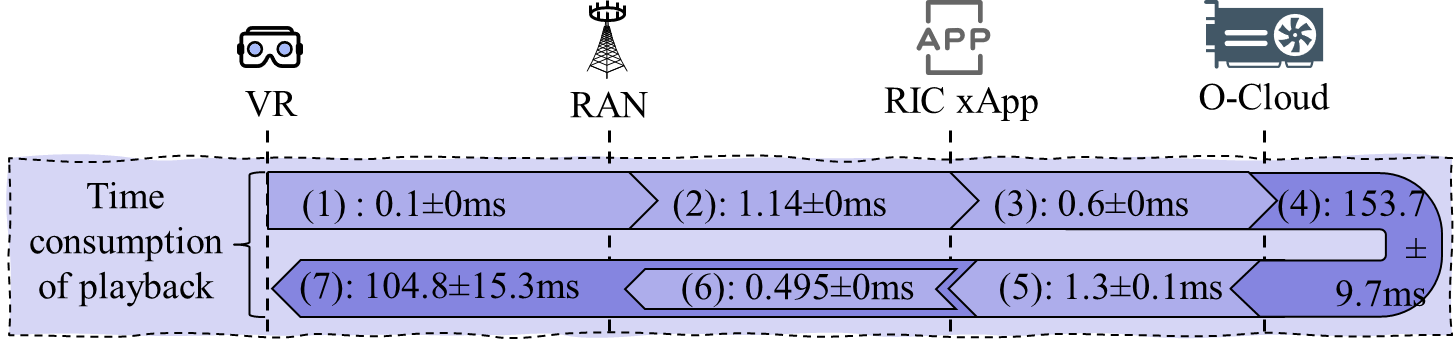}
  } \\
  \subfigure[Decision-making time consumption.\label{fig:deploy_alg_time_consumption}]{
    \includegraphics[width=0.95\linewidth]{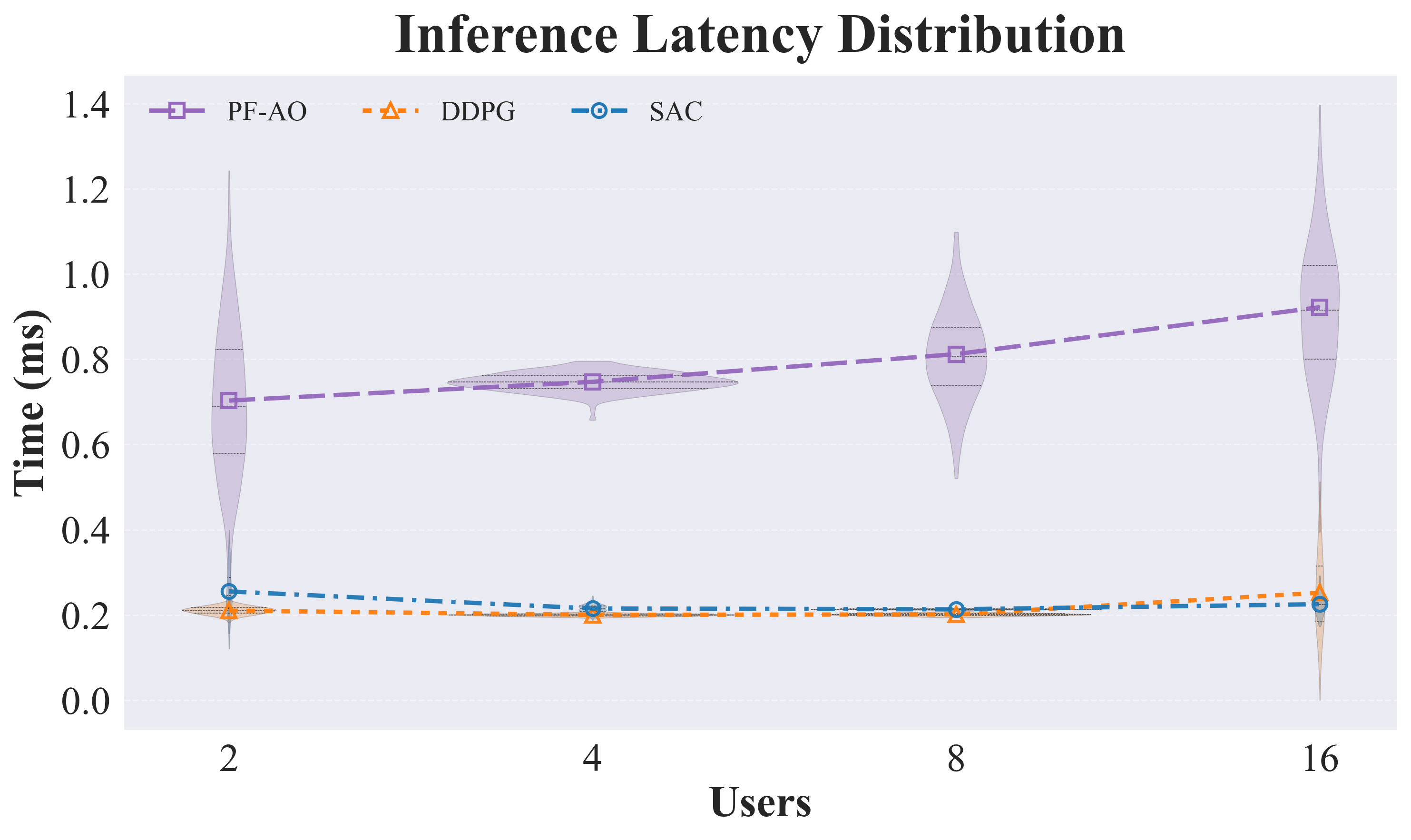}
  }
  \caption{Time consumption of the XR O-RAN communication system.}
  \label{fig:deploy_vr_alg_time_consumption}
\end{figure}

Fig.~\ref{fig:deploy_vr_time_consumption} reports the per-frame time breakdown of the playback pipeline, with steps (1-7) aligned with Fig.~\ref{fig:deploy_vr_oran} and the sequence in Fig.~\ref{fig:imvid_reconstruction_and_playback}. The process consists of VR-status upload from the HMD to the gNB via the CPE, frame rendering on the O-Cloud, and frame download from the gNB to the HMD. Rendering dominates the overall latency, followed by download, while upload is minimal due to the small payload, indicating that compute-intensive rendering is the primary bottleneck in our setup.

Fig.~\ref{fig:deploy_vr_time_consumption} summarizes the end-to-end dataflow latency from the HMD through the gNB, RIC, and O-Cloud and back, following the same order as Fig.~\ref{fig:imvid_reconstruction_and_playback}. We report latency as mean $\pm$ standard deviation to reflect both average overhead and variability. With 10 samples at $1080\text{P}$, the average per-frame latency is $262.7\,\text{ms}$ (SD $= 14.33\,\text{ms}$), where the main components are HMD-to-gNB upload $0.1\,\text{ms}$ (SD $= 0\,\text{ms}$), gNB decision $1.14\,\text{ms}$ (SD $= 0\,\text{ms}$), RIC-to-O-Cloud $0.6\,\text{ms}$ (SD $= 0\,\text{ms}$), O-Cloud rendering $153.7\,\text{ms}$ (SD $= 9.7\,\text{ms}$), O-Cloud-to-gNB $1.3\,\text{ms}$ (SD $= 0.1\,\text{ms}$), RIC-to-gNB control $0.495\,\text{ms}$ (SD $= 0\,\text{ms}$), and gNB-to-HMD download $104.8\,\text{ms}$ (SD $= 15.3\,\text{ms}$).

Fig.~\ref{fig:deploy_alg_time_consumption} compares inference time for SAC, DDPG, and PF-AO, excluding the average-allocation baseline since it has no nontrivial inference. After a 10-run warm-up, we measure each method 10 times on the testbed and report mean and standard deviation, with violin plots showing the distributions. Under eight users, the average inference time is 0.213 ms for SAC, 0.201 ms for DDPG, and 0.812 ms for PF-AO, where SAC is close to DDPG and PF-AO is slower and more variable due to its iterative procedure. Inference time increases with user count, while SAC and DDPG grow mildly, supporting real-time decision making in multi-user settings and meeting per-frame latency requirements.

\subsection{Traffic and Resolution Analysis}

We deployed the video generation service on the O-Cloud to serve XR users. The O-Cloud host runs an XR xApp that controls RAN resource allocation and compute scheduling. The prototype therefore provides XR users with coordinated resource allocation, power control, and video generation services.

\begin{figure}[!htb]
    \centering
    \includegraphics[width=1\linewidth]{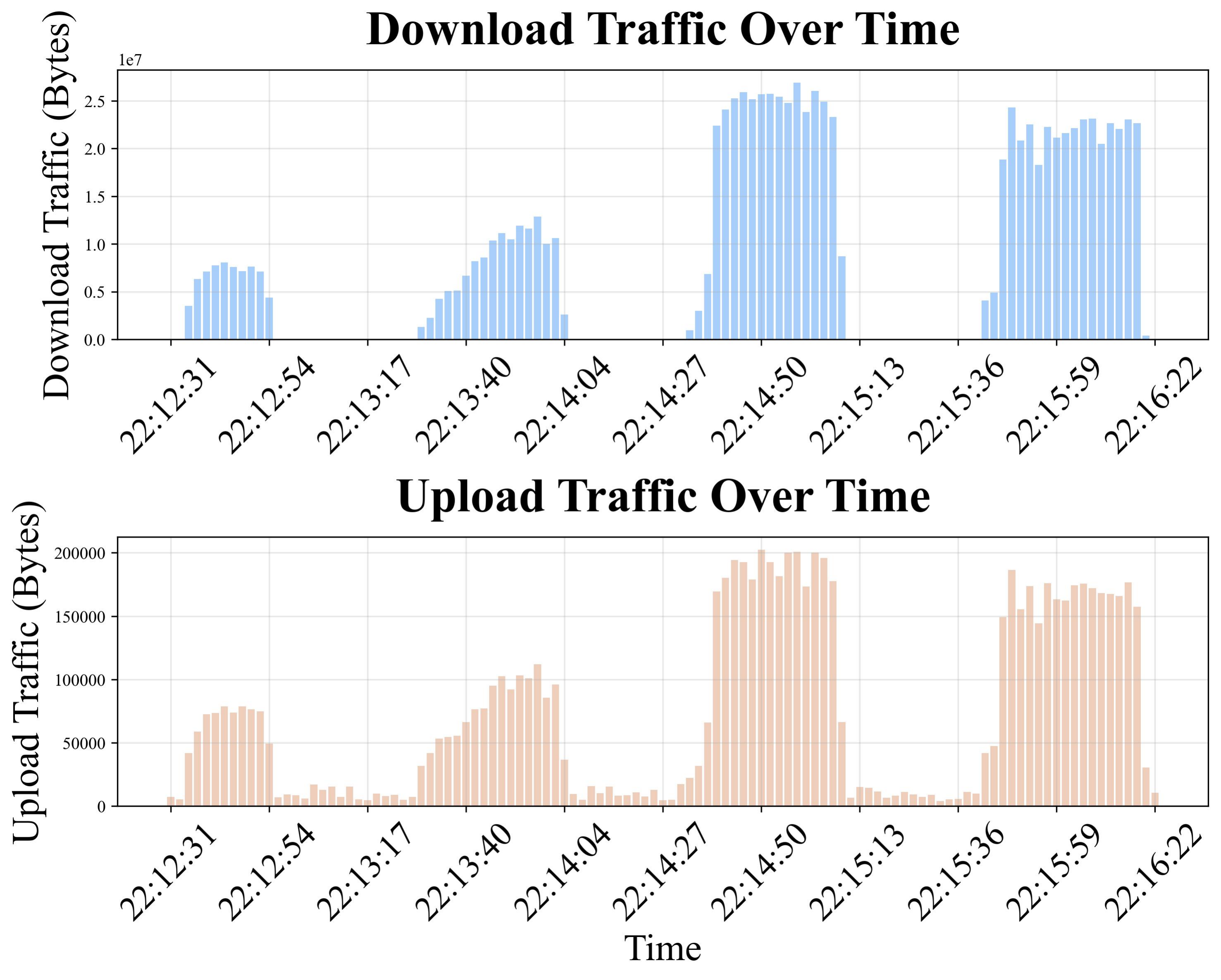}
  \caption{Measured download and upload traffic at the gNB during resolution changes.}
  \label{fig:deploy_traffic}
\end{figure}
Fig.~\ref{fig:deploy_traffic} presents real traffic measurements collected at the gNB every five seconds. 
The horizontal axis indicates different time points, and the XR device plays video at different resolutions. The video resolutions are $270\times480$, $480\times640$, $720\times1280$, and $1080\times1920$, with each resolution lasting approximately 30 seconds.
The upload traffic primarily consists of VR pose data uploaded by the HMD, which is relatively small and remains below 0.2~MB/s. The download traffic mainly comprises video data transmitted from the gNB to the HMD, which is significantly larger, typically ranging from 8~MB/s to over 25~MB/s.

Both upload and download traffic at the gNB are generally stable, with no major fluctuations observed. As the user selects different video resolutions, the download traffic changes accordingly-the higher the resolution, the greater the download throughput. However, when the resolution reaches $1080\times1920$, the download traffic does not continue to increase; instead, it decreases slightly. This is because each video frame at this resolution is very large, reaching up to 5.94 MB (without compression) per frame. The increased rendering and transmission time on the server side leads to a reduction in video frame rate, which in turn lowers the download throughput. As a result, users experience lower network speeds when viewing $1080\times1920$ video.

\begin{figure}[!htb]
    \centering
    \includegraphics[width=1\linewidth]{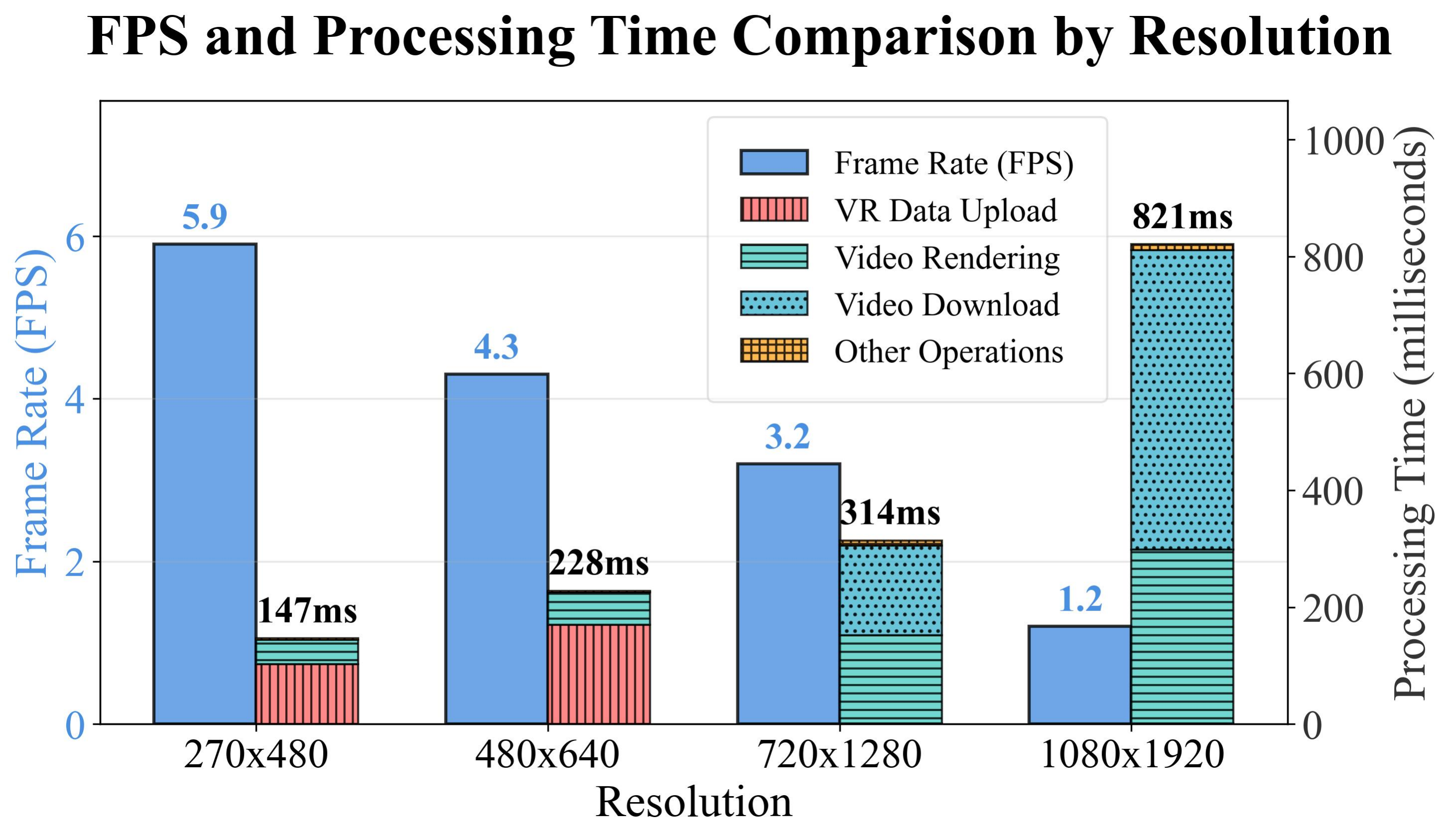}
  \caption{Deployment of Video Generation Service}
  \label{fig:deploy_fps_delay}
\end{figure}
Fig.~\ref{fig:deploy_fps_delay} illustrates the deployment of the video rendering service and the relationship between frames per second (FPS) and time cost at different resolutions. The video generation service is deployed on the O-Cloud server. Users connect to the gNB via the 5G network, and the gNB forwards user requests to the XR xApp and the O-Cloud. After rendering the video, the server transmits the generated video frame back to the gNB, which then delivers it to the user.

As shown in Fig.~\ref{fig:deploy_fps_delay}, the FPS decreases and the time cost increases as the resolution rises. At lower resolutions (the first two cases), the time cost for uploading VR status to the XR xApp is relatively high. 
This is because the download speed is fast; after image delivery, frequent but small VR status updates incur long queuing times, increasing upload delay. 
When the resolution increases (the latter two cases), the time cost for video rendering becomes the dominant factor affecting video generation latency. Moreover, as the resolution continues to rise, the rendering time cost also increases, which negatively impacts the user's immersive experience.

\section{Conclusion}

We presented an O-RAN-assisted immersive volumetric video system that unifies XR rendering and RAN control. We designed an ImViD playback architecture, implemented a 5G O-RAN prototype with open-source components, and formulated a joint communication-compute resources allocation problem. A DRL-based solution using SAC improves QoE, latency, and fairness over non-learning baselines in simulation, while prototype measurements corroborate the latency-throughput tradeoffs across resolutions. 

Future work will consider large-scale deployments, online learning with safety constraints, and integration with emerging 6G primitives.
Edge computing and digital twins open several promising directions for extending the proposed O-RAN-based architecture, where the Non-RT/Near-RT/RT RIC hierarchy coordinates cloud–edge–device compute and RAN control for scalable XR delivery. Potential avenues include distributed or edge-partitioned RIC/O-Cloud deployments with safe online learning for robustness and fault isolation, two-timescale orchestration that couples long- and short-term decisions, and richer QoE optimization that integrates eye-tracking/viewport prediction with energy awareness, while enabling digital-twin interaction enhanced by generative AI under perceptual QoE models.

\section*{Acknowledgment}
The paper was partly funded by Jiangsu Major Project on Fundamental Research (Grant No.: BK20243059), Natural Science Foundation of China (Grant No. 62531008, 62132004), Gusu Innovation Project (Grant No.: ZXL2024360), High-Tech District of Suzhou City (Grant No.: RC2025001).

\bibliography{./CiteBib}
\bibliographystyle{IEEEtran}


\section*{Biographies}
\textbf{Yao Wen} (Graduate Student Member, IEEE) received a bachelor's degree from Nanjing Forestry University, Nanjing, China, in 2021, and a master degree from China University of Mining and Technology, Xuzhou, China, in 2024. He is pursuing a doctoral degree at Nanjing University (Suzhou Campus). His research interests include wireless communications, federated learning, edge computing, open radio access network (O-RAN) and AI-RAN.

\textbf{Luping Xiang} (Senior Member, IEEE) received the B.Eng. degree (Hons.) from Xiamen University, China, in 2015, and the Ph.D. degree from the University of Southampton, in 2020. From 2020 to 2021 He was a Research Fellow with the Next Generation Wireless Group, University of Southampton. In November 2021, he joined the University of Electronic Science and Technology of China (UESTC) as a faculty member, and in September 2024, he joined Nanjing University as an Assistant Professor. His main research areas include native intelligence at wireless communication, end-to-end transmission technology, computer vision, and integrated sensing and communication transmission.

\textbf{Kun Yang} (Fellow, IEEE) received his PhD from the Department of Electronic \& Electrical Engineering of University College London (UCL), UK. He is currently a Chair Professor of Nanjing University and an affiliated professor at the University of Essex. His main research interests include wireless networks and communications, communication-computing cooperation, and AI (artificial intelligence) for wireless. He has published 600+ papers and filed 50 patents. He serves on the editorial boards of a few IEEE journals (e.g., IEEE WCM, TVT, TNB). He is a Deputy Editor-in-Chief of IET Smart Cities Journal. He has been a Judge of GSMA GLOMO Award at World Mobile Congress – Barcelona since 2019. He was a Distinguished Lecturer of IEEE ComSoc, a Recipient of the 2024 IET Achievement Medals and the Recipient of 2024 IEEE CommSoft TC’s Technical Achievement Award. He is a Member of Academia Europaea (MAE), a Fellow of IEEE, a Fellow of IET and a Distinguished Member of ACM.

\end{document}